\documentclass[fleqn,usenatbib]{mnras}

\usepackage{newtxtext,newtxmath}

\usepackage[T1]{fontenc}

\DeclareRobustCommand{\VAN}[3]{#2}
\let\VANthebibliography\thebibliography
\def\thebibliography{\DeclareRobustCommand{\VAN}[3]{##3}\VANthebibliography}

\usepackage{graphicx}	
\usepackage{amsmath}	
\RequirePackage{mcode}
\usepackage{listings}
\usepackage{booktabs}
\usepackage[export]{adjustbox}

\title[S+FP: SExtractor detection of nearby galaxies in large surveys]{The S-PLUS Fornax Project (S+FP): SExtractor detection and measurement of nearby galaxies in large photometric surveys}
\author[R. F. Haack et al.]{R. F. Haack,$^{1,2}$\thanks{E-mail: rodrihaack@fcaglp.unlp.edu.ar}
A. V. Smith Castelli,$^{1,2}$
C. Mendes de Oliveira,$^{3}$
F. Almeida-Fernandes,$^{3}$
F. R. Faifer,$^{1,2}$
\newauthor
A.R. Lopes,$^{1}$
Y. Jaffe,$^{4,5}$
R. Demarco,$^{6}$
C. Lima-Dias,$^{7,8}$
L. Lomelí-Nuñez,$^{9}$
G. P. Montaguth,$^{7}$
\newauthor
W. Schoenell,$^{10}$
T. Ribeiro,$^{11}$
A. Kanaan$^{12}$\\
\\
$^{1}$Instituto de Astrofísica de La Plata, UNLP-CONICET, Paseo del Bosque s/n, La Plata, B1900FWA, Argentina\\\
$^{2}$Facultad de Ciencias Astrónomicas y Geofísicas, Universidad Nacional de La Plata, Paseo del Bosque s/n, La Plata, B1900FWA, Argentina\\
$^{3}$Universidade de São Paulo, IAG, Rua do Matão 1226, Sao Paulo, SP, Brazil\\
$^{4}$Departamento de Física, Universidad Técnica Federico Santa María, Av. España, 1680 Valparaíso, Chile\\
$^{5}$Instituto de Física y Astronomia, Universidad de Valparaíso, 1111 Gran Bretana, Valparaíso, Chile\\
$^{6}$Institute of Astrophysics, Facultad de Ciencias Exactas, Universidad Andrés Bello, Sede Concepción, Talcahuano, Chile\\
$^{7}$Departamento de Astronomía, Universidad de La Serena, Av. Raúl Bitrán 1305, La Serena, Chile\\
$^{8}$Instituto Multidisciplinario de Investigación y Postgrado, Universidad de La Serena, Raúl Bitrán 1305, La Serena, Chile\\
$^{9}$Valongo Observatory, Federal University of Rio de Janeiro, Ladeira Pedro Antonio 43, Saude Rio de Janeiro, RJ, 20080-090, Brazil\\
$^{10}$GMTO Corporation 465 N. Halstead Street, Suite 250 Pasadena, CA 91107, USA \\
$^{11}$Rubin Observatory Project Office, 950 N. Cherry Ave., Tucson, AZ 85719, USA \\
$^{12}$Departamento de Física - CFM - Universidade Federal de Santa Catarina, PO BOx 476, 88040-900, Florianópolis, SC, Brazil\\
}

\date{Accepted XXX. Received YYY; in original form ZZZ}

\pubyear{2023}

\begin{document}
\label{firstpage}
\pagerange{\pageref{firstpage}--\pageref{lastpage}}
\maketitle

\begin{abstract}
All-sky multi-band photometric surveys represent a unique opportunity of exploring rich nearby galaxy clusters up to several virial radii, reaching the filament regions where pre-processing is expected to occur. These projects aim to tackle a large number of astrophysical topics, encompassing both the galactic and extragalactic fields. In that sense, generating large catalogues with homogeneous photometry for both resolved and unresolved sources that might be interesting to achieve specific goals, imposes a compromise when choosing the set of parameters to automatically detect and measure such a plethora of objects. In this work we present the acquired experience on studying the galaxy content of the Fornax cluster using large catalogues obtained by the Southern Photometric Local Universe Survey (S-PLUS). We realized that some Fornax bright galaxies are missed in the S-PLUS iDR4 catalogues. In addition, Fornax star-forming galaxies are included as multiple detections due to over-deblending. To solve those issues, we performed specific SExtractor runs to identify the proper set of parameters to recover as many Fornax galaxies as possible with confident photometry and avoiding duplications. From that process, we obtained new catalogs containing 12-band improved photometry for $\sim3\times10^6$ resolved and unresolved sources in an area of $\sim208$ deg$^2$ in the direction of the Fornax cluster. Together with identifying the main difficulties to carry out the study of nearby groups and clusters of galaxies using S-PLUS catalogs, we also share possible solutions to face issues that seem to be common to other ongoing photometric surveys. 
\end{abstract}

\begin{keywords}
galaxies: clusters -- surveys -- techniques: photometric 
\end{keywords}



\section{Introduction}

In recent decades, large astronomical surveys employing imaging in three or more photometric bands have gained significant prominence, causing a profound transformation in various scientific domains, including the study of large-scale cosmic structures and the evolution of galaxies. With the advent of the Sloan Digital Sky Survey (SDSS; \citealp{York2000}), studies that previously relied on the analysis of several hundreds or a few thousands of galaxies, now can be tackled with millions of them. In order to allow the analysis of such volumes of imaged objects with homogeneous information, it is needed to automatically determine, among others, photometric and structural parameters for all of them. 

A decade ago, the morphological classification of galaxies was based on visual inspection of their images by specialists (\citealp{Nair2010,Ann2015}) or by citizens not professionally trained in astronomy \citep{GZ1,GZ2,Simmons2017}. In recent decades, numerical algorithms have been used to determine the main morphological parameters of galaxies \citep{Spiekermann1992,Storrie-Lombardi1992,Walmsley2020}. However, the impressive data volumes expected for ongoing surveys like the Legacy Survey of Space and Time (LSST, 17 TB per dark night; \citealp{Tyson2002,Axelrod2006}) carried out by the Vera C. Observatory, 
will transform  into unfeasible all those methods. 

The automatization of the photometric measurements for a large amount of images using specific software like \texttt{SExtractor} \citep{SExtractor} arises as the natural solution in the above context. However, it enforces a compromise on the settings in order to apply the same process to different types of objects. In the local Universe, specifically, galaxies can display large projected sizes in the sky imposing an added challenge in homogenizing the photometry.

In this paper we present the analysis of such a compromise in the case of the obtention of the photometric catalogues from the Southern Photometric Local Universe Survey \citep[S-PLUS,][]{Mendes2019} at the distance of the Fornax galaxy cluster ($D~\sim$ 20 Mpc). 
S-PLUS is imaging $\sim$9,300 deg$^2$ of the celestial sphere in twelve optical bands using a 0.8-m robotic telescope (the T80-South) at the Cerro Tololo Inter-American Observatory, in Chile. The survey consists of four main zones, including two non-contiguous ones at high Galactic latitudes (|b| > 30º, 8,000 deg$^2$) and two zones of the Galactic plane and bulge (for an additional 1,300 deg$^2$). S-PLUS uses the Javalambre 12-band magnitude system \citep{J-PLUS} that includes the 5 broadband filters u, g, r, i, z and 7 narrowband filters to cover some specific spectral lines. Although there are several studies currently underway and planned for the future based on wide-field imaging in the Southern Celestial Hemisphere, S-PLUS provides a unique sampling of the optical spectrum thanks to its seven narrow-band filters. Four public datasets have already been released and as of today, 70\% of what was initially planned has been mapped. In particular, the fourth internal data release (iDR4) available to the S-PLUS collaboration includes 1,629 fields covering 3,000 deg$^2$ of the southern sky.  For more details on S-PLUS, we refer the reader to \citet{Mendes2019} and \citet{Almeida-Fernandes2022}.

Counting with upgraded photometric catalogues offers not only the possibility of studying, for example, fainter sources than those included in the released catalogues, but also the chance to significantly improve several ongoing studies initially based on the released data. Such a  methodology becomes particularly relevant in nearby clusters, where very extended galaxies can be found, thus enhancing the photometry and source detection in regions such as Fornax \citep[][hereafter, Paper\,I]{PaperI}, Hydra \citep{Lima-Dias2021}, and Antlia. In addition, it can help unfolding techniques to better analyze galaxy morphologies (\citealt{Lima-Dias2024,Bom2021,Buzzo2022,Montaguth2023}) and it can facilitate the analysis of stellar population properties and emission lines within galaxies \citep{Thaina-Batista2023}. Furthermore, an improved source detection capability may play a crucial role in the identification of various structures in the local Universe, including distant galaxy clusters \citep{Werner2023}, and may provide better estimations of photometric redshifts \citep{Lima2022}. Also, the classification of stars, quasars, and galaxies \citep{Nakazono2021,Costa-Duarte2019} may benefit from it as well as systematic studies of the stellar populations of low surface brightness (LSB) galaxies \citep{Barbosa2020}, the analysis of compact objects such as globular clusters (GCs) \citep{Hartmann2022,Buzzo2022}, and even the search for planetary nebulae (PNe) \citep{Gutierrez-Soto2020}. 

The paper is organized as follows. In Section\,\ref{sec:data} we present the initial data and the motivation of this work. In Section\,\ref{sec:methodology}, we present the methodology used to tackle the identified issues and, in Section\,\ref{sec:results}, the results. In Section\,\ref{sec:improvements} we present an improvement of our method in order to solve remaining problems as well as several tests on galaxies belonging to more distant clusters and on images of other photometric surveys. In Section\,\ref{sec:discussion} we present a discussion of our results and, in Section\,\ref{sec:conclusions}, the summary and conclusions. In this paper we work with AB magnitudes \citep{Oke1974ApJS...27...21O} and we assume a distance modulus of $(m-M)=31.51$ for Fornax \citep{Blakeslee2009}. At the distance of Fornax, 1 arcsec subtends 0.1 kpc.

\section{Initial Data and Motivation}
\label{sec:data}

\begin{figure}
\centering
\includegraphics[width=1.0\columnwidth]{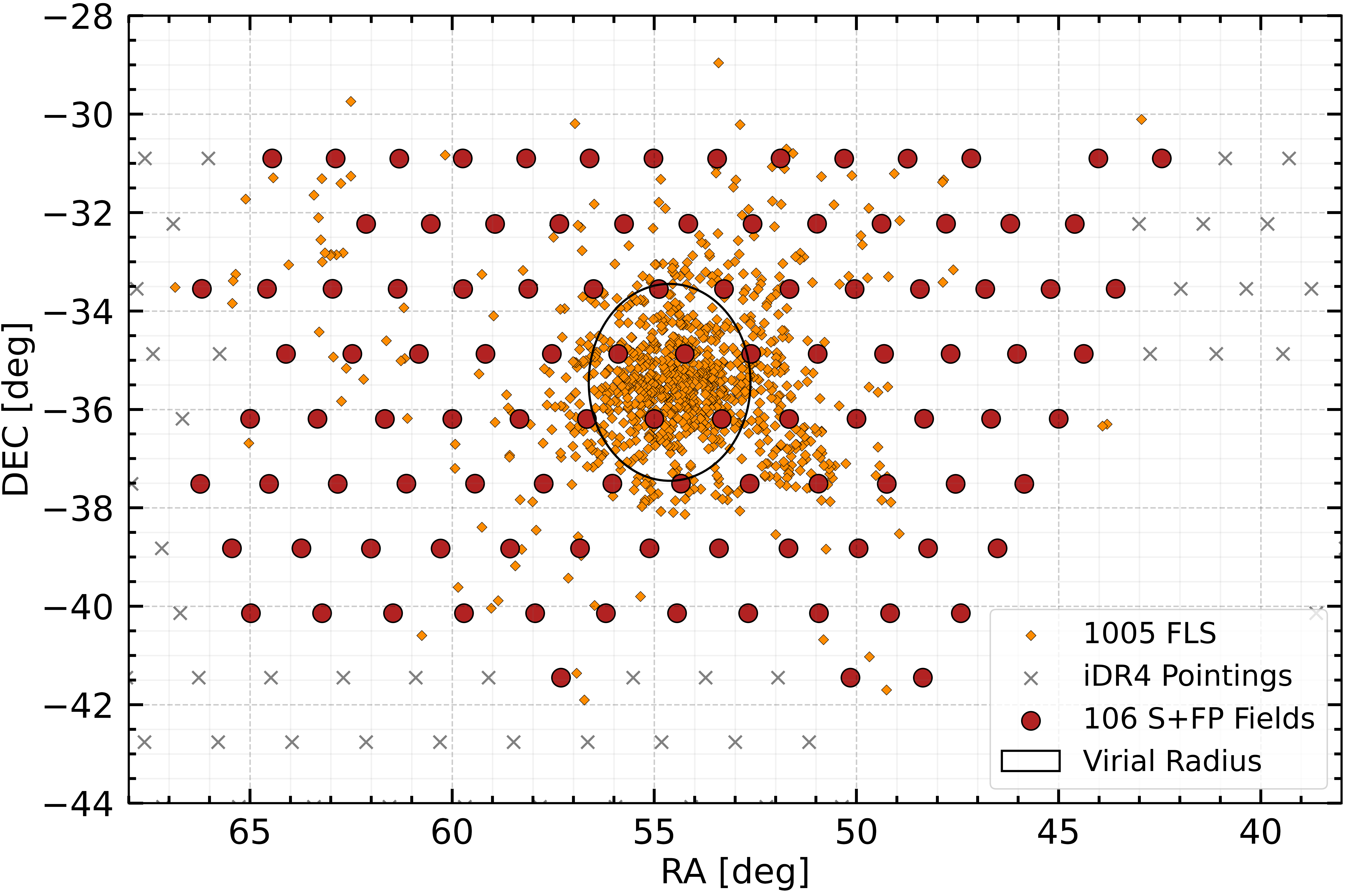}
 \caption{Spatial distribution of the 106 S+FP fields and the Fornax Literature Sample (FLS) introduced in Paper\,I.} 
 \label{fig:pointings}
\end{figure}

The S-PLUS Fornax Project (S+FP; Paper\,I) aims at performing an extensive analysis of the Fornax galaxy cluster using the images corresponding to 106 S-PLUS pointings observed as part of iDR4 (see Figure \ref{fig:pointings}). These 106 pointings cover $\sim$ 208 deg$^2$ centered around NGC\,1399, the brightest galaxy of the Fornax cluster, and 1000 galaxies included in 21 Fornax galaxy catalogues from the literature that we will refer to hereafter as the Fornax Literature Sample (FLS; see Paper\,I). This implies that in the context of the S+FP, $12\times106 = 1,272$ images are being analyzed. Each of them has a resolution of 0.55 arcsec pixel$^{-1}$ and a total size\footnote{Each 11,000 $\times$ 11,000 pixel$^2$ image corresponds to the co-adding of multiple observations (usually three) for each filter. The main goal of this process is to increase the saturation limit of the CCD. During this process, the observations are dithered to mitigate the influence of bad pixels, which has the side effect of slightly increasing the field of view of the observations.} of 11,000 $\times$ 11,000 pixel$^2$, which translates to a field of view (FoV) of $\sim 1.4 \times 1.4$ deg$^2$. There are catalogues with homogeneous photometry obtained as part of the iDR4 of S-PLUS for all the S+FP pointings. Those catalogues have been obtained from \texttt{SExtractor} \citep{SExtractor} runs using a common set of parameters to all iDR4 fields that are not optimized to detect any particular type of object. In that sense, the set of parameters used to obtain the iDR4 catalogues represent a compromise for detecting both resolved and unresolved sources that could be interesting to tackle galactic or extragalactic scientific goals. The magnitudes included in those catalogues are calibrated to the standard AB system and they are not corrected for interstellar extinction. The astrometric precision of the iDR4 catalogues is $\Delta \alpha$ = 0.010 arcsec and $\Delta \delta$ = 0.027 arcsec. For more details on the photometry of the iDR4 catalogues, we refer the reader to \citet{Almeida-Fernandes2022}. 

\begin{figure}
\centering
\includegraphics[width=1.0\columnwidth]{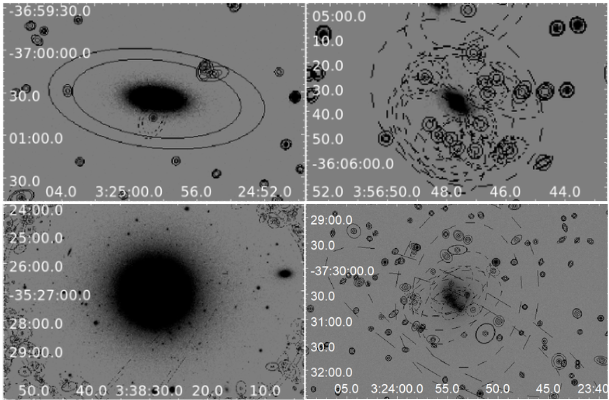}
 \caption{Four examples of aperture images corresponding to the S-PLUS iDR4 \texttt{SExtractor} parameters. The top-left image shows a galaxy displaying neither substructures nor a nearby bright companion on which a good detection is achieved. The bottom-left image shows NGC\,1399 and how the neighboring objects are not detected. The top- and bottom-right images show two star-forming galaxies excesively deblended. } 
 \label{fig:problematica}
\end{figure}

After making a cross-match between the iDR4 catalogues of the 106 S+FP fields and the FLS, we noticed that several Fornax galaxies were not included in those catalogues. Therefore, in order to understand why those objects were missed, we decided to make some \texttt{SExtractor} runs on specific S+FP fields using the same set of parameters of iDR4. Through the \texttt{SExtractor's} aperture images obtained from those runs, we realized that some of the missing objects were not properly separated from a bright neighbouring galaxy. In addition, from the aperture images of our test, we noticed that star-forming galaxies tend to be excesively deblended and, therefore, a proper cross-match between the FLS and the iDR4 catalogues become difficult (see Figure\,\ref{fig:problematica} for some examples). As a consequence, we decided to try to identify sets of \texttt{SExtractor} parameters that are more suitable to properly detect, in an automatic way, galaxies at the Fornax cluster distance.

\section{Methodology}
\label{sec:methodology}
\subsection{Automation Code}
\label{subsec:code}

One of the complexities of this work is that, for each observed field, there are 12 images, each corresponding to one of the S-PLUS filters. Thus, for each field, 12 input configuration files must be edited in order to be able to run \texttt{SExtractor} on the individual images. This implies that, prior to the specific work with \texttt{SExtractor}, a first stage of programming in \texttt{Python} is required to automate the editing of all  ($106\times12 = 1,272$) configuration files. This is done in order for \texttt{SExtractor} to take into account specific quantities of each image such as gain, saturation and seeing. This is necessary since the images provided by S-PLUS are constructed from different numbers of exposures that are taken with different exposure times, depending on the filter. On the other hand, the observing conditions may vary over the course of an observing run.

Also, running \texttt{SExtractor} in dual mode involves, in our case, an additional set of images called "detection" images. These images are usually generated from the sum of several of the broad-band images or by considering an average of them. In our case, and following the procedure used in the iDR4 of S-PLUS, the detection image is the sum of the g, r, i and z images of each field. The main motivation for constructing a sum image is to increase the signal-to-noise ratio (S/N) in order to reliably detect and measure as many faint objects as posible. Thus, there is a second stage prior to the actual execution of \texttt{SExtractor} that consists in the construction of such detection images for the 106 fields which we plan to work on.

Once the detection images are obtained and all the configuration files are edited, we proceed to the execution of \texttt{SExtractor} in dual mode on each of the 106 S+FP fields. As a result of this execution we obtain, for each field, a set of control images (aperture images in our particular case) that show the quality of the detection of the different objects. We also obtain 12 catalogues, one for each S-PLUS filter, containing the photometric information for all the detected objects in the field. Thus, after the execution of \texttt{SExtractor}, additional programming is required to concatenate the individual catalogues which provides a master catalogue containing the complete photometry in the 12 S-PLUS filters of all the objects detected in a given field.

\begin{figure}
\centering
\includegraphics[width=1.0\columnwidth]{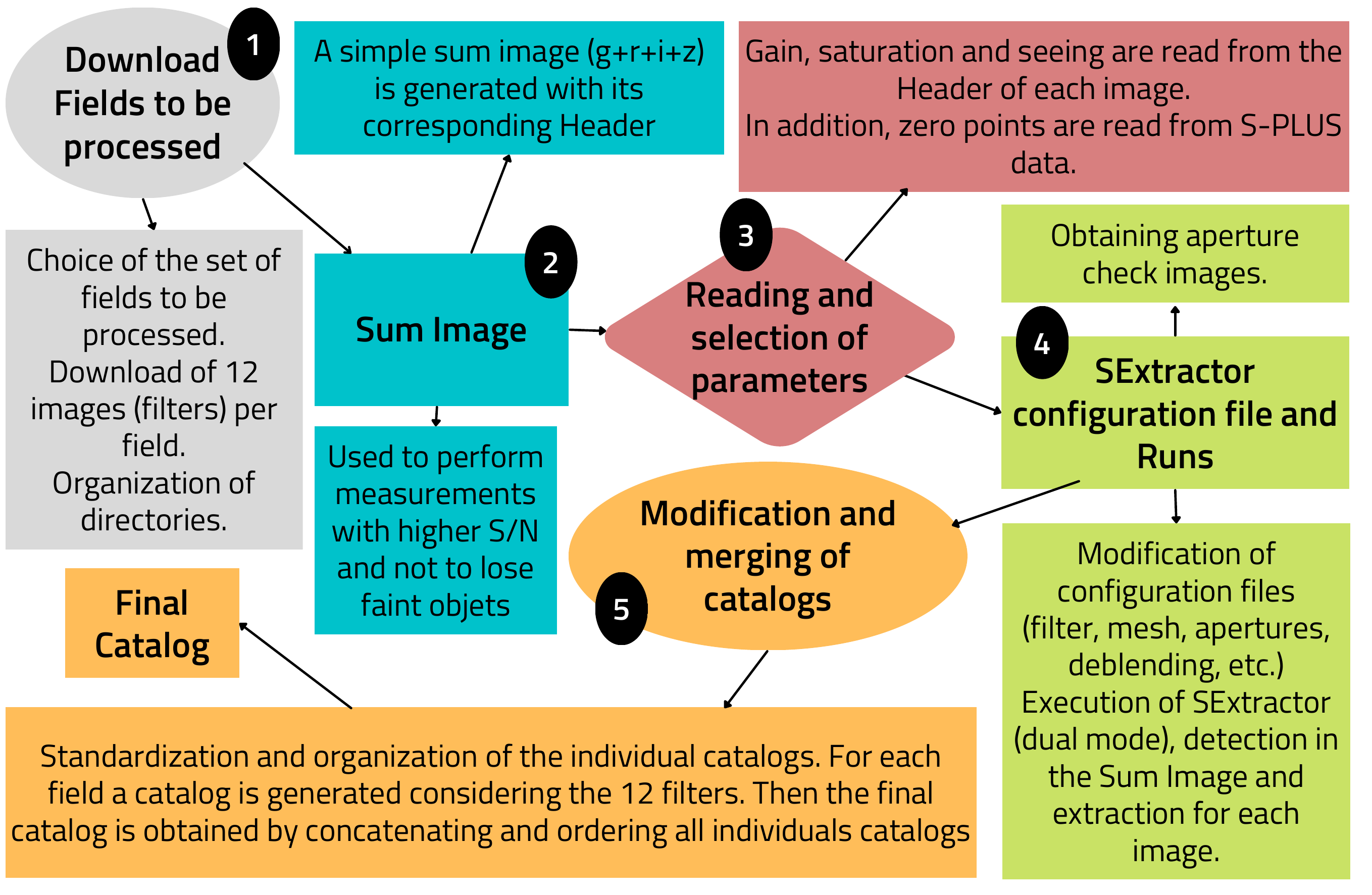}
 \caption{Basic scheme of the five steps in the automation process.} 
 \label{fig:cuadro}
\end{figure}

It is worth mentioning that performing a complete \texttt{SExtractor} run may lead to unsatisfactory results, which will imply modifying the parameters of the configuration files in order to obtain a better result. Thus, the procedure of editing the configuration files and running \texttt{SExtractor} may be repeated several times. In this sense, as a final product in terms of programming, we have generated a code in \texttt{Python} that integrates all the previously mentioned stages, which is called  \texttt{Measurement and Extraction with Sextractor on Surveys Images (MESSI\footnote{\small
\url{https://github.com/rodrihaack/MESSI}})}. The input data of the code are the general configuration file to be edited and the set of images which to work on. A schematic plot of the five steps of the automation process is shown in Figure\,\ref{fig:cuadro}. 
They can be summarized as follows: 

\begin{enumerate}
    \item[1-] Downloading the images through the \texttt{splusdata} package\footnote{\url{https://github.com/Schwarzam/splusdata}}.
    \item[2-] Building the detection (sum) images.
    \item[3-] Building the configuration files by reading the zero points for each field and each filter from \texttt{splus.cloud}\footnote{\url{https://splus.cloud}}, and saturation, gain and seeing values from the headers of the images.
    \item[4-] Running \texttt{SExtractor}.
    \item[5-] Concatenation of the catalogues obtained for each field in each filter. 
\end{enumerate}

\subsection{S+FP SExtractor runs: RUN\,1 and RUN\,2}
After numerous tests with different parameter sets, two of them were chosen to be systematically repeated. We named those general configuration files as RUN\,1 and RUN\,2. The RUN\,1 parameters are intended to detect faint galaxies and compact objects located in the vicinity of bright galaxies, while those of RUN\,2 aim at obtaining a good characterization of bright and extended galaxies. Therefore, compact objects, such as ultra-compact dwarfs (UCDs) or GCs in the vicinity of Fornax galaxies, would be better detected with RUN\,1, while brighter and more extended Fornax objects would be better characterized with RUN\,2. The parameters in which S-PLUS iDR4, RUN\,1 and RUN\,2 differ are shown in Table\,\ref{tab:Parameters}. 

As it can be seen from the table, in the parameters involved in the detection and extraction of the objects, an important difference between RUN\,1 and RUN\,2 is the value of {\footnotesize DETECT\_MINAREA}. According to those values, RUN\,2 will only detect objects displaying 10 or more connected pixels, while RUN\,1 will do it as long as the number of connected pixels is larger than 5. For the detection of objects above a given threshold, there is the option of applying a filter. In the case of RUN\,1, a small pyramid function is set by default, and corresponds to a $3\times3$ "all-ground" convolution mask with FWHM$=$2 pixels, while RUN\,2 uses a Gaussian filter, imposing a $9\times9$ convolution mask of a Gaussian PSF with FWHM$=$5.0 pixels. This will favor RUN\,2 to detect bright and extended objects over RUN\,1.

The parameters {\footnotesize DEBLEND\_NTHRESH} and {\footnotesize DEBLEND\_MINCONT} are linked to the process of {\it deblending} or separating objects. That is, through those parameters, \texttt{SExtractor} makes the decision whether a group of adjacent pixels above {\footnotesize DETECT\_THRESH} correspond to a single object or constitute the superposition of several ones. In that sense, the values given to those parameters in Table\,\ref{tab:Parameters} allow RUN\,1 to detect close and faint objects, while RUN\,2 seeks to ensure that bright objects with resolved structure are not fragmented as separate objects. 

\begin{figure*}
 \includegraphics[width=2.0\columnwidth]{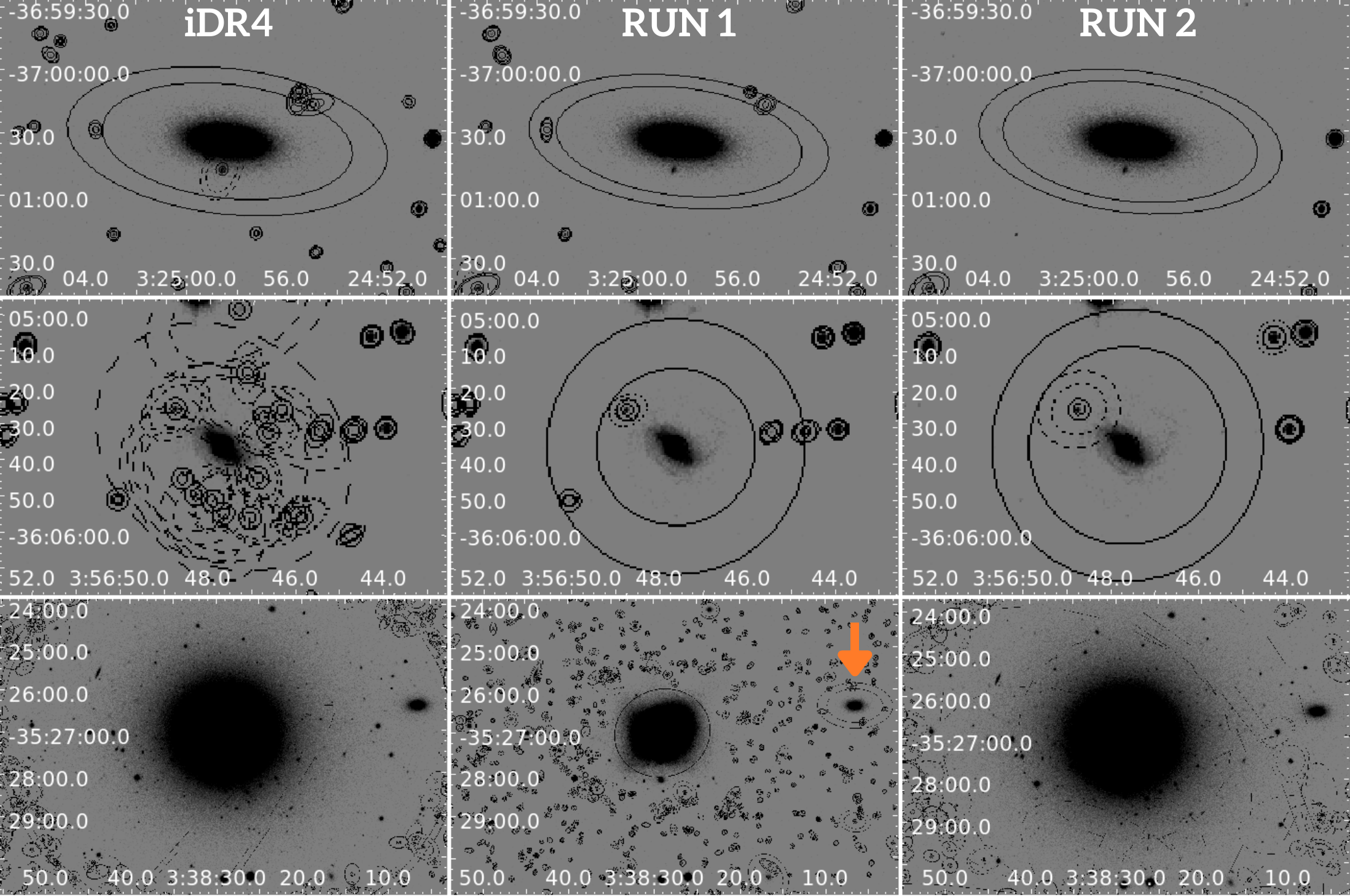}
 \caption{{\it Left:} Aperture images obtained using \texttt{SExtractor} input parameters of S-PLUS iDR4. {\it Center:} Aperture images obtained using RUN\,1 \texttt{SExtractor} input parameters. The orange arrow in the bottom panel points to a dwarf galaxy detected with this run and missed with the input parameters of S-PLUS iDR4. {\it Right:} Aperture images obtained from RUN\,2 \texttt{SExtractor} input parameters. All panels in a row show the same object.}
 \label{fig:comparacion}
\end{figure*}

\begin{figure}[!h]
\centering
\includegraphics[width=1.0\columnwidth]{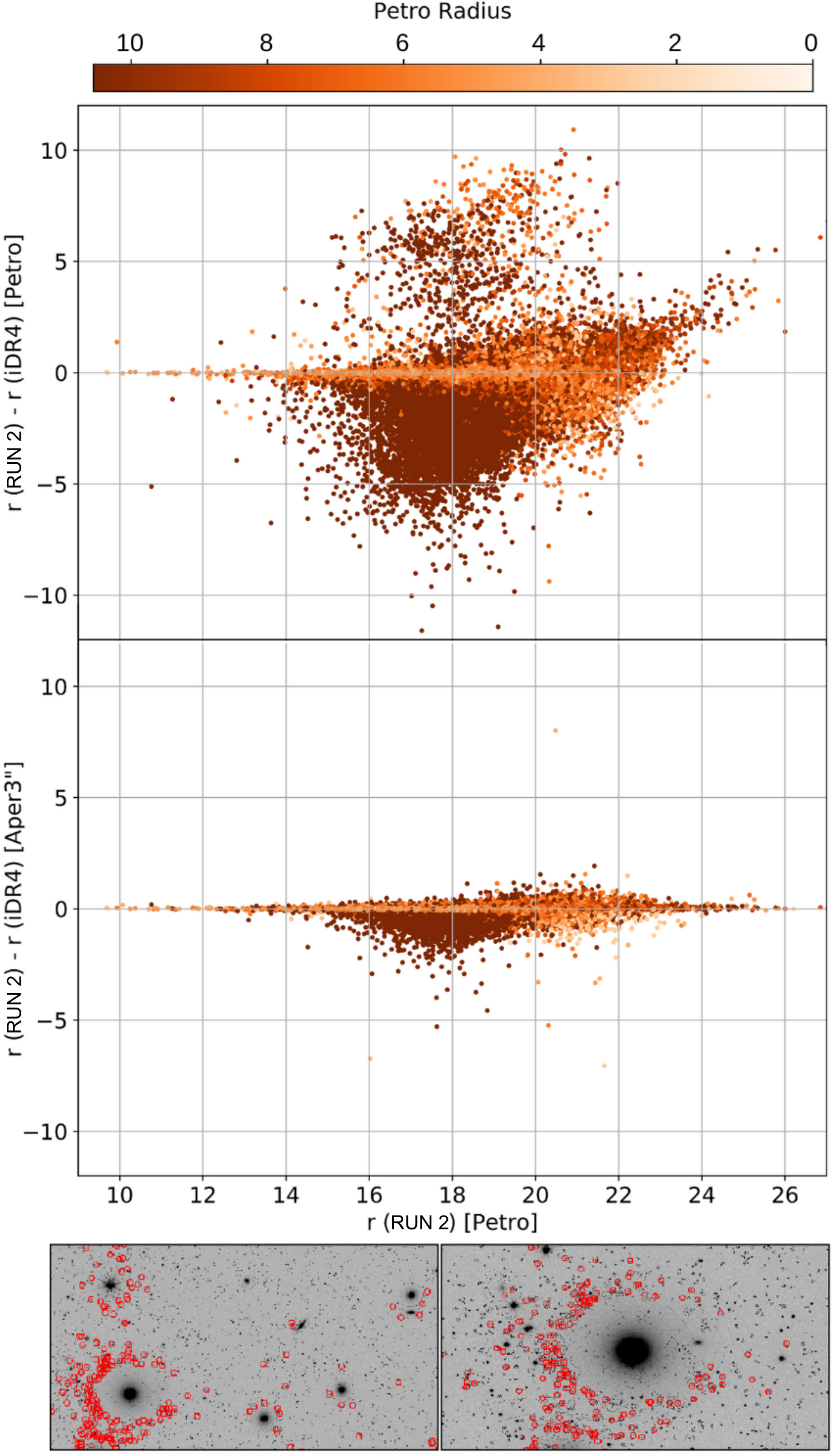}
 \caption{Comparison of magnitudes obtained with iDR4 and RUN\,2. The x-axis in the upper and central panels represents the Petrosian magnitude in the r-band obtained with the RUN\,2 dataset. In the upper panel, the y-axis showcases the discrepancies in Petrosian magnitudes in the r-band measured by RUN\,2 and iDR4, while the central panel provides a similar comparison using 3 arcsec aperture magnitudes. The shared color bar corresponds to the PETRO$\_$RADIUS in the r-band measured with RUN\,2, providing insights into the extent of the objects. In the upper panel, an homogeneous brown cloud is discernible toward negative values (RUN\,2 magnitudes brighter than those obtained by iDR4), encompassing two distinct categories of objects. Some of them are small objects positioned near highly luminous and extended counterparts, highlighted by red circles in the two images in the lower panel. Others are brighter and more extended objects. The latter category is the subset that persists in the brown cloud seen in the central panel. The magnitude differences in smaller objects adjacent to brighter galaxies are less pronounced when considering 3 arcsec apertures.} 
 \label{fig:petro_radius}
\end{figure}

The substantial differences in the background estimation of both RUNs have to do with the fact that RUN\,1 performs a local calculation with a grid subregion size according to the frequently used values, while RUN\,2 applies a global estimation, manually setting the value of the background to 0. That is, RUN\,2 does not perform a background estimation at all. These differences are linked to the idea that RUN\,1 is expected to detect faint and compact objects, while RUN\,2 aims at lowering the effect of deblending in order to properly detect and measure large objects. It should be noted that the S-PLUS images are already sky subtracted.

\begin{table}
\centering
\caption{Comparison of the input parameters of \texttt{SExtractor} used to obtain the catalogues of iDR4 \textit{(left)}, RUN\,1 \textit{(centre)} and RUN\,2 \textit{(right)}}
\label{tab:Parameters}
\begin{tabular}{|l|l|l|l|}
\textbf{Parameters} & \textbf{iDR4} & \textbf{RUN\,1} & \textbf{RUN\,2} \\
\hline
 DETECT\_MINAREA & 4 & 5 & 10 \\ 
 DETECT\_THRESH & 1.1 & 1.5 & 1.5 \\  
 ANALYSIS\_THRESH & 3. & 1.5 & 1.5 \\
 FILTER & Y & Y & Y \\ 
 FILTER\_NAME & tophat\_3\_3x3 & default & gauss\_5\_9x9 \\  
 DEBLEND\_NTHRESH & 64 & 32 & 64 \\
 DEBLEND\_MINCONT & 0.0002 & 0.005 & 0.001 \\ 
 BACK\_SIZE & 256 & 64 & 524 \\  
 BACK\_FILTERSIZE & 7 & 3 & 3 \\
 BACKPHOTO\_TYPE & LOCAL & LOCAL & GLOBAL \\
 BACKPHOTO\_THICK & 48 & - & - \\
 BACK\_TYPE & - & - & MANUAL \\
 BACK\_VALUE & - & - & 0.0 \\
 \hline
\end{tabular}
\end{table}

Figure\,\ref{fig:comparacion} shows nine \texttt{SExtractor} aperture images. Each row corresponds to the same region in the sky. The left column images were obtained with the parameters adopted by S-PLUS iDR4, the center column frames, with those of RUN\,1, and the right column images come from RUN\,2. The top row is a clear example in which iDR4 works quite well and RUN\,1 or RUN\,2 do not represent a significant improvement. In the center row, it is noticeable how iDR4 fails to properly detect the whole galaxy as a single object, while RUN\,1 and RUN\,2 do it correctly. In the bottom row, it is interesting to see how RUN\,1 properly detects a dwarf galaxy in the vicinity of a bright giant galaxy, which is marked with an orange arrow in the central panel.

\section{Results}
\label{sec:results}

\subsection{Catalogues}
After performing both runs in the 106 S+FP fields, two catalogues were obtained containing homogeneous photometry of resolved and unresolved objects, all located in a total area of the southern sky of $\sim$208 deg$^2$ in the direction of the Fornax galaxy cluster. The RUN\,1 catalogue includes 2,900,926 objects, and the RUN\,2 catalogue, 1,390,237 objects.

In order to perform a quantitative analysis of the obtained catalogues, Figure\,\ref{fig:petro_radius} shows a comparison between several types of magnitudes measured with the iDR4 set of parameters and those of RUN\,2, which is expected to properly detect the most extended galaxies. The x-axis in both panels corresponds to the Petrosian magnitude in the r-band obtained with RUN\,2. The y-axis in the top panel shows the difference between the Petrosian magnitudes in the r-band measured by RUN\,2 and iDR4. The y-axis in the central panel shows a similar difference but using magnitudes measured within apertures of 3 arcsec diameter. The color bar, common to both panels, represents the PETRO$\_$RADIUS in the r-band obtained with RUN\,2, which gives an idea of how extended the objects are. 

It can be seen that in the upper panel there is an homogeneous brown cloud towards negative values. That is, that cloud includes extended obects that are brighter in the RUN\,2 catalogue. These objects correspond to two different types: relatively small objects close to very bright and extended galaxies (indicated with red circles on the two images in the bottom panel), and very bright and extended objects. The latter are the objects defining the much smaller brown cloud that persists in the central panel. The brightness of small objects close to bright galaxies does not differ significatively when considering 3 arcsec apertures. In summary, iDR4 and RUN\,2 Petrosian magnitudes display a significant difference for extended objects and for objects close to bright sources. However iDR4 aperture photometry works well for small objects away from bright sources. A brief analysis of the photometric depth of the catalogues can be found in Appendix \ref{sec:apen2}.

\subsection{Photometry Validation}
\label{subsec:validation}

It is necessary to compare with other surveys to validate the photometry obtained by RUN\,1 and RUN\,2.  To do that, both catalogues were constrained using the following conditions:

\begin{itemize}
    \item\footnotesize{CLASS\_STAR\_{\it g}} < 0.35 \& \footnotesize{CLASS\_STAR\_{\it r}} < 0.35 \& \footnotesize{CLASS\_STAR\_{\it i}} < 0.35,

    \item \footnotesize{{\it g}\_AUTO} < 22 mag \&     \footnotesize{{\it r}\_AUTO} < 22 mag \& \footnotesize{{\it i}\_AUTO} < 22 mag.
\end{itemize}

\noindent The first criterion allows us to select resolved objects with a good degree of confidence of being galaxies, avoiding unresolved objects with a high probability of being stars. The second criterion allows us to select objects with expected brightnesses for galaxies taking into account the depth of our data. It will also discard objects not well measured by \texttt{SExtractor} that will display a value of 99 in those magnitudes. This results in a restricted RUN\,1 catalogue containing 350,078 objects (12\% of the sources included in the original catalogue), 
and a restricted RUN\,2 catalogue containing 357,574 objects (26\% of the sources included in the original catalogue). 
We will refer to these catalogues as RUN\,1$_{restr}$ and RUN\,2$_{restr}$, respectively.
After this process, the RUN\,2$_{restr}$ catalogue was taken as a reference since, for extended objects included in both the RUN\,1$_{restr}$ and RUN\,2$_{restr}$ catalogues, the latter will contain the most reliable APER and PETRO magnitudes.  
Next, using coordinates, the objects included in the RUN\,1$_{restr}$ catalogue, that were not found in RUN\,2$_{restr}$, were selected to be added to RUN\,2$_{restr}$. 
In that way, we obtain a RUN\,1+2 master catalogue.

\begin{figure}
    \centering \includegraphics[width=1.0\columnwidth]{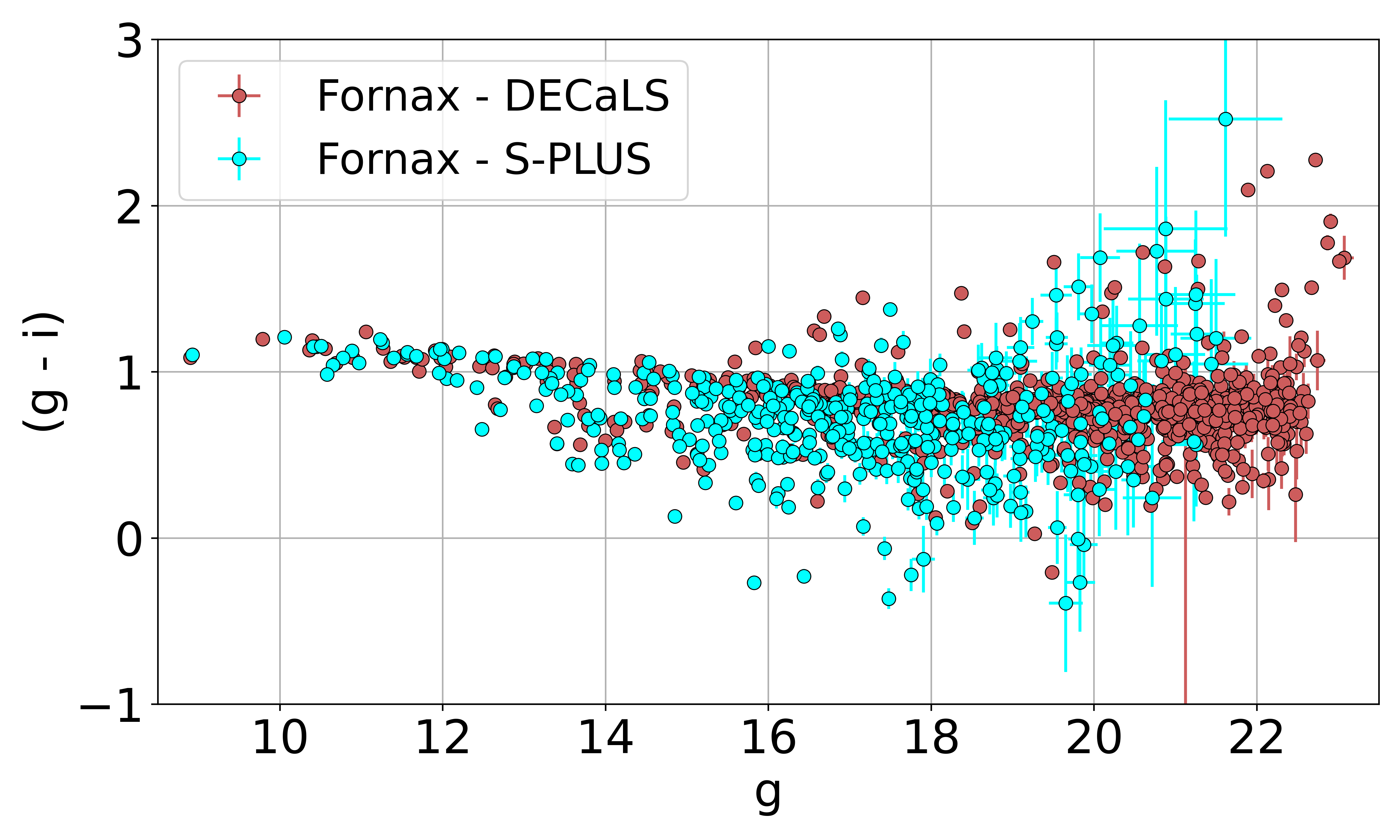}
    \caption{(g-i) vs. g diagram from the RUN 1+2 photometry (cyan dots) also showing the same galaxies but with photometry provided by the DECaLS DR10 (red dots). 
    We can see that the broad-band photometry obtained from RUN\,1+2 recovers the colour-magnitude relation of the Fornax cluster in a consistent manner to that obtained with the DECaLS photometry. However, RUN\,1+2 mangitudes and colors display larger errors than those of DECaLS because the latter is a deeper survey than S-PLUS. 
    }
    \label{CMR_LS}
\end{figure}

In order to asses the quality of the photometry obtained from RUN\,1 and RUN\,2, a crossmatch of the RUN\,1+2 catalogue with the FLS was performed to recover cluster members that were detected by \texttt{SExtractor} with some degree of confidence in the S+FP images. From that cross-match, a sample of 443 Fornax galaxies was recovered. 
In Figure\,\ref{CMR_LS} we show a color-magnitude plot of the galaxies detected in Fornax (cyan dots). We can see that the color-magnitude relation (CMR) of Fornax is well recovered with our photometry. For comparison, we overimpose the colors and magnitudes of the same set of galaxies now obtained from the DECam Legacy Surveys (DECaLS; \citealp{Dey2019}) (red dots). We can see the good agreement between both photometries, although RUN\,1+2 data display larger photometric errors towards the faint end of the CMR. This is expected since DECaLS is a deeper survey than S-PLUS. 

\begin{figure}
\centering
\includegraphics[width=1.0\columnwidth]{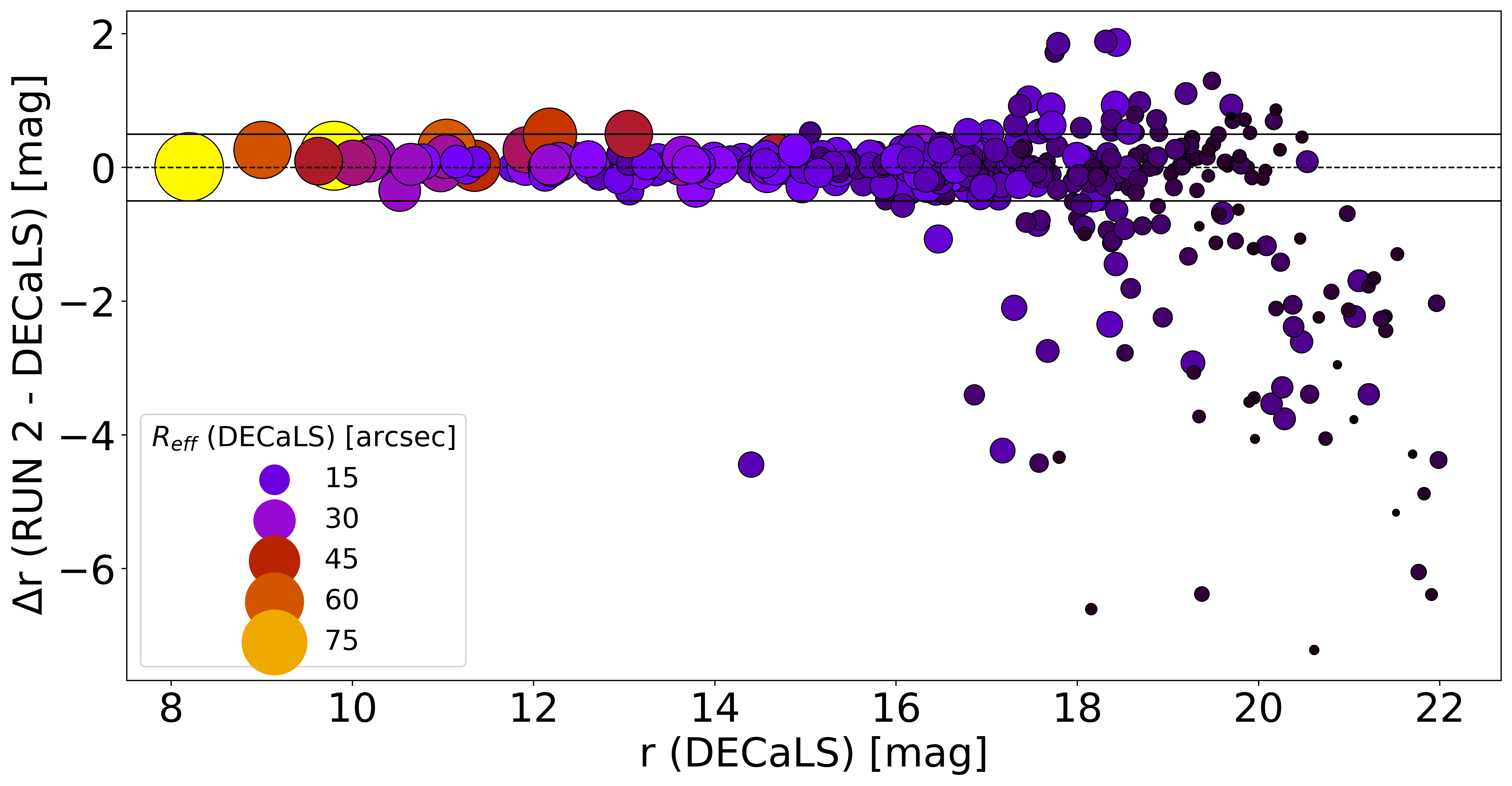}
 \caption{Difference between DECaLS and RUN\,2 r-band magnitude versus DECaLS r-band magnitudes for objects detected by both DECaLS and RUN\,2. The horizontal lines on the graph correspond to $\Delta r=\pm0.5$ mag and $\Delta r=0$ mag. The sizes of the data points are directly proportional to the effective radius derived from DECaLS.} 
 \label{fig:Delta_LS_Run2}
\end{figure}

\begin{figure}
\centering
\includegraphics[width=1.0\columnwidth]{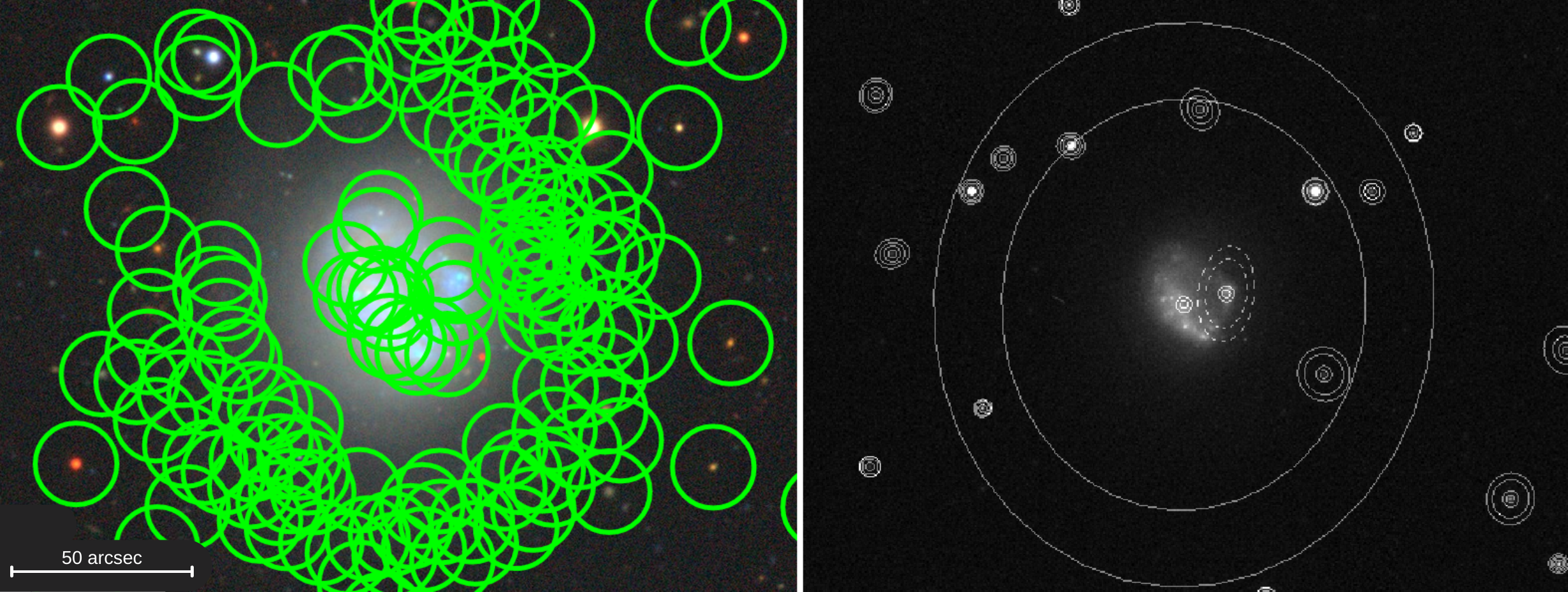}
 \caption{Example of 
 an over-deblendend galaxy by DECaLS. In the left panel, the green circles indicate the multiple detections obtained by DECaLS on the image of ESO\,301-11. In the right panel we show the RUN\,2 detections on the same galaxy.} 
 \label{fig:Deblending_LS_Run2}
\end{figure}

In order to quantify the agreement between the photometry of extended galaxies obtained by our \texttt{SExtractor} runs and DECaLS, Figure\,\ref{fig:Delta_LS_Run2} shows a plot of $\Delta$r vs. r, where the y-axis represents the difference between the r-total magnitudes of the objects detected by both DECaLS and RUN\,2. In this analysis we will focus on RUN\,2 as it is the run that better characterizes the total magnitude of large galaxies. The horizontal lines correspond to  $\Delta r=\pm$ 0.5 mag and  $\Delta r=$ 0 mag. The size of the dots is proportional to the effective radius provided by DECaLS. We choose a value of 0.5 mag as it is usually adopted as a flexible photometric error. Towards the bright end of the sample the difference is very small, while for objects near the faint end, some of them fall outside the range -0.5 mag < $\Delta$r < 0.5 mag. By inspecting these objects separately, we realized that they are objects that are over-deblended in DECaLS. An example of such objects is shown in Figure\,\ref{fig:Deblending_LS_Run2}, that compares the number of detections obtained by DECaLS (right panel) and RUN\,2 (left panel). Such an excessive deblending in the case of DECaLS implies that the catalogues will not contain photometric information of the object as a single source, but it will contain fragmented information. As a consequence, the DECaLS magnitude of that particular source will be underestimated. 
This exercise allowed us to realize that over-deblending might be a common issue to several surveys. 

\section{Improvements and Additional Tests}
\label{sec:improvements}
\subsection{Over-deblending of large galaxies with complex structures: RUN\,3}
\label{subsec:run3}

Despite RUN\,1 and RUN\,2 improve the detection and measurement of star-forming galaxies, faint sources and compact objects near bright galaxies, they still fail in obtaining a good characterization of the most extended galaxies displaying a complex internal structure. An illustrative case is NGC\,1365, the brightest spiral galaxy of the Fornax cluster. To solve this issue, we have tested a third set of parameters that we call RUN\,3. Table \ref{tab:RUN3} shows the SExtractor parameters adopted by RUN\,3 that differ from RUN\,1 and RUN\,2. The main characteristics of this new run are: 
\begin{itemize}
    \item A minimum of 3,000 connected pixels for detection, ensuring the detection of very extended objects;
    \item high sensitivity levels for detection and analysis, thus guaranteeing the detection and measurement of bright objects, leaving aside the weak ones; 
    \item 16 deblending thresholds are considered, being this the lowest value allowed;
    \item the contrast parameter is set high enough to only allow deblending of close objects similar in brightness; 
    \item background estimation is not performed as in RUN\,2.
\end{itemize}

\begin{table}
\centering
\caption{\texttt{SExtractor} parameters adopted by RUN\,3 that differ from RUN\,1 and RUN\,2.}
\label{tab:RUN3}
\begin{tabular}{|l|r|}
\textbf{Parameters } & \textbf{RUN\,3} \\
\hline
DETECT\_MINAREA & 3000 \\
DETECT\_THRESH & 5.0 \\
ANALYSIS\_THRESH & 5.0 \\
FILTER & Y \\
FILTER\_NAME & gauss\_5.0\_9x9.conv \\
DEBLEND\_NTHRESH & 16 \\
DEBLEND\_MINCONT & 0.1 \\
BACK\_SIZE & 4500 \\
BACK\_FILTERSIZE & 3 \\
BACKPHOTO\_TYPE & GLOBAL \\
BACK\_TYPE & MANUAL \\
BACK\_VALUE & 0.0 \\
\hline
\end{tabular}
\end{table}

\begin{figure}
\centering
\includegraphics[width=1.0\columnwidth]{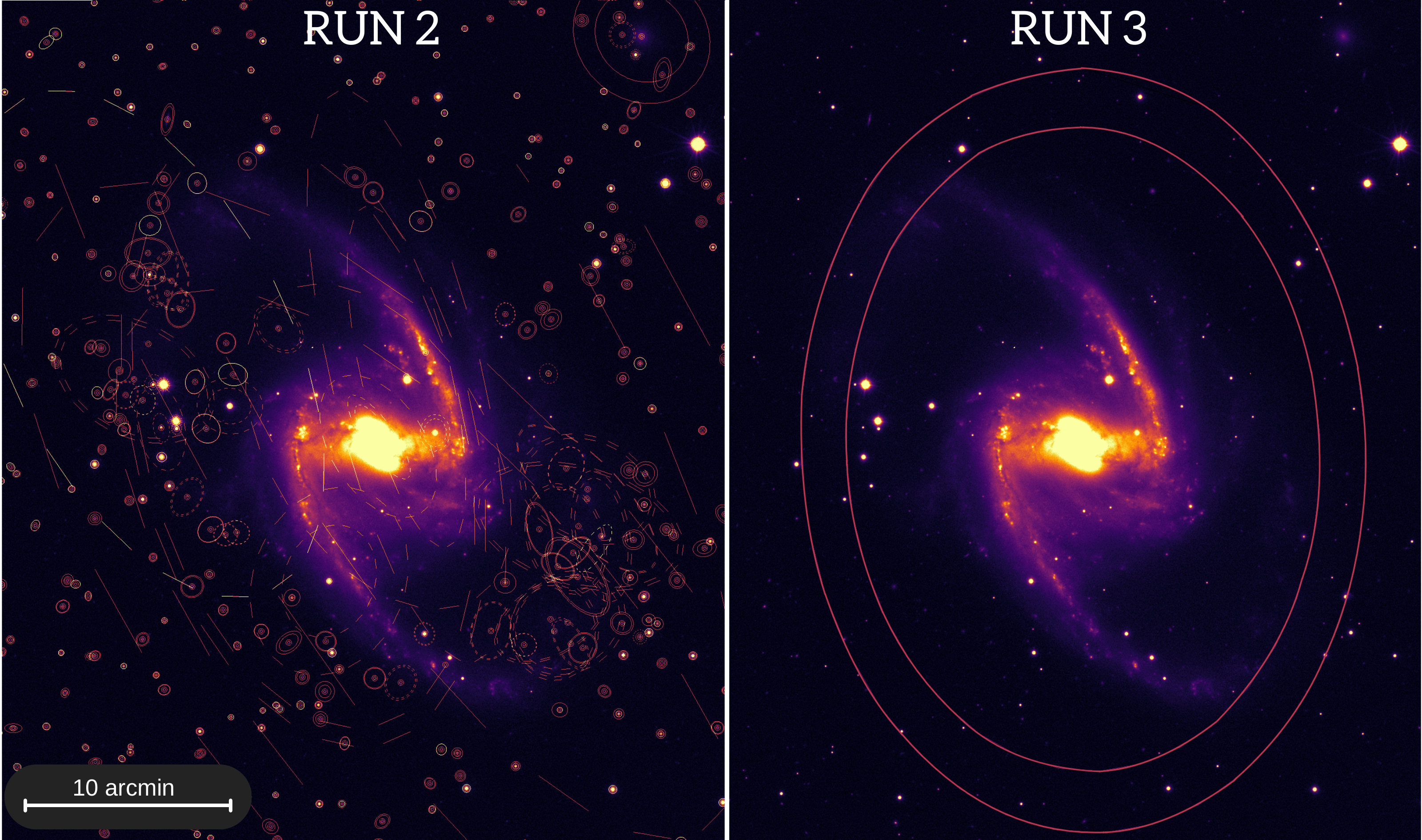}
 \caption{The left image corresponds to the detections performed by RUN\,2 on the galaxy NGC\,1365, a bright spiral in the Fornax cluster. The right image shows the detection as a single source (not deblended) performed by RUN\,3.} 
 \label{fig:RUN 3 examples}
\end{figure}

\noindent In the left panel of Figure\,\ref{fig:RUN 3 examples}, we show the detections made by RUN\,2 on NGC\,1365, and in the right panel, the detections made by RUN\,3 as a single source. 

\begin{figure*}
\centering
\includegraphics[width=1.0\textwidth]{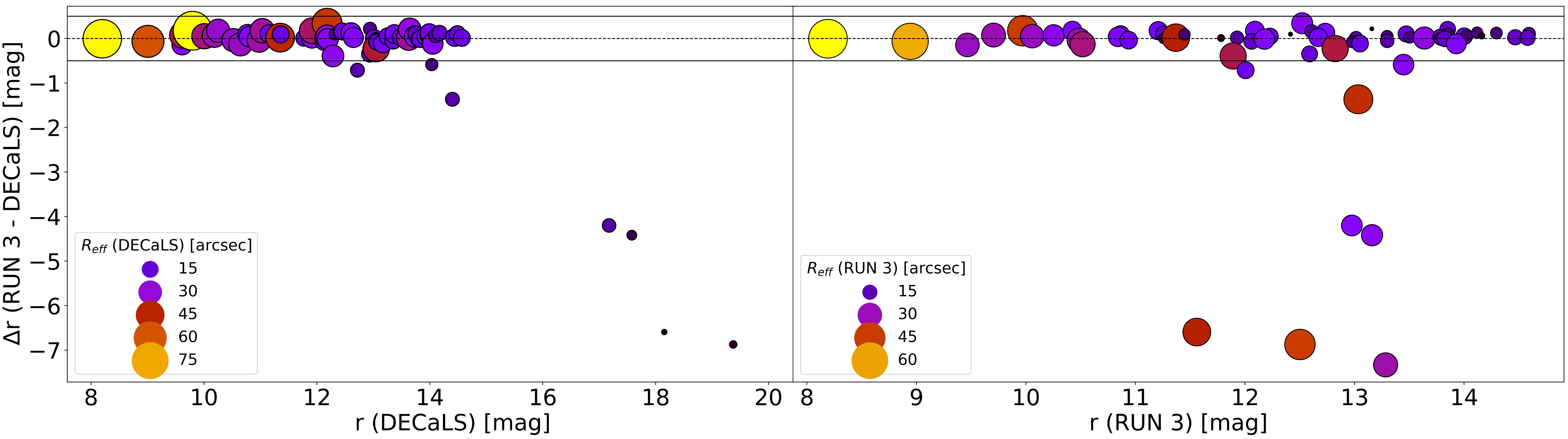}
 \caption{Diagram depicting the relationship $\Delta$r vs. r where the vertical axis represents the difference in r-band magnitude between the DECaLS and the RUN\,3 photometry. The horizontal axis represents the magnitude measured by DECaLS (left panel) and RUN\,3 (right panel). The horizontal lines on the graph correspond to deviations of $\Delta r=\pm0.5$ mag and $\Delta r=0$ mag. The point sizes is proportional to the effective radius derived by DECaLS and RUN\,3 respectively.} 
 \label{fig:RUN3-LEGACY}
\end{figure*}

\begin{figure}
\centering
\includegraphics[width=1.0\columnwidth]{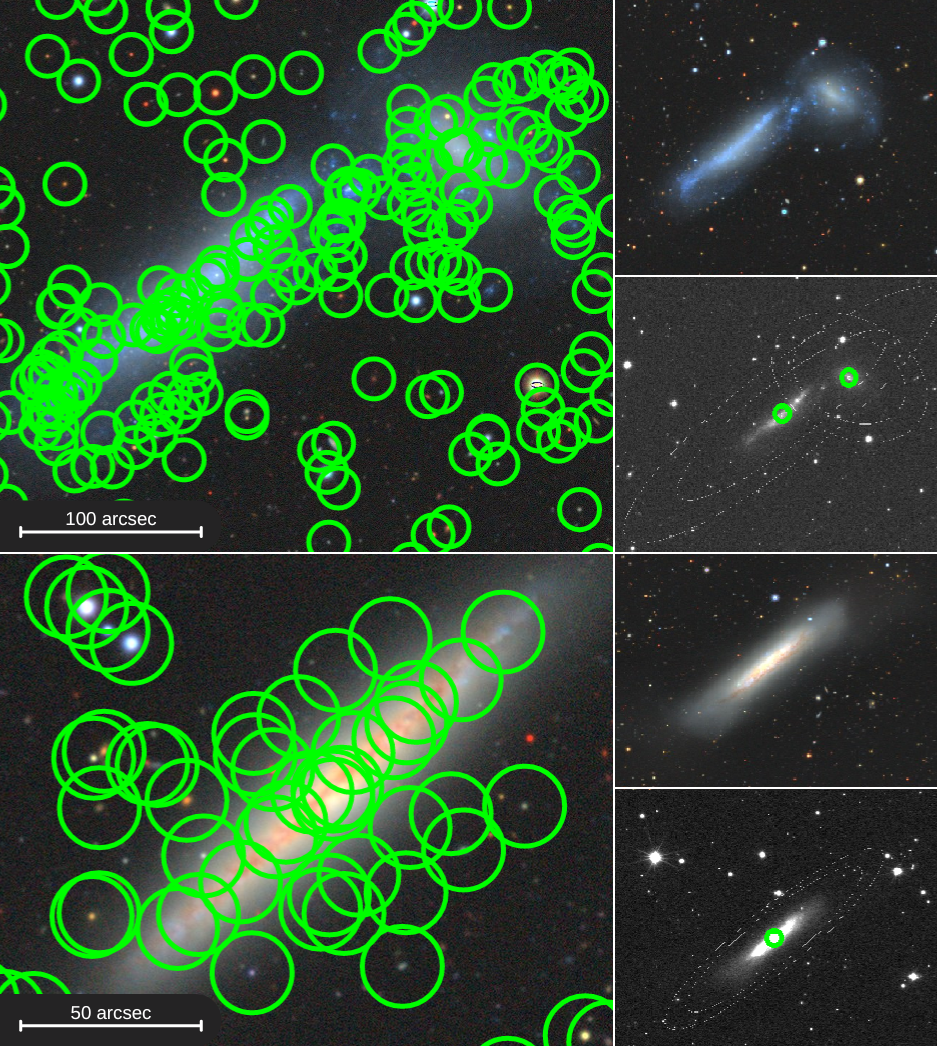}
 \caption{Two additional examples of objects displaying complex structures that are not well characterized by RUN\,1 and RUN\,2. In this case, we show the multiple DECaLS detections on these galaxies, which emphasizes the fact that the over-deblending issue is also present in other photometric surveys. For each example, in the upper right panel, the RGB image of DECaLS without the detections and, in the lower right panel, we show the apertures image of RUN\,3 when executed on the DECaLS frames.}
 \label{fig:LS_examples_Run7}
\end{figure}

 In Figure\,\ref{fig:RUN3-LEGACY} we present a similar plot to Figure\,\ref{fig:Delta_LS_Run2}, taking into account the same objects, but now considering the differences in magnitudes between RUN\,3 and DECaLS. On the x-axis of the left panel, we show the magnitude in the r-band measured by DECaLS, while on the x-axis of the right panel, the one measured by RUN\,3. Both panels share the vertical axis and the color bar indicates the effective radius given by DECaLS and RUN\,3, respectively. In the left panel it is noticeable how several objects display remarkable differences in the magnitudes measured by RUN\,3 and DECaLS. By inspecting these sources separately, we found that they are bright galaxies excessively deblended for which brighter magnitudes than those measured by DECaLS are expected. Two examples can be seen in Figure\,\ref{fig:LS_examples_Run7}. For each example, in the left panel we show the multiple DECaLS detections overimpossed on the RGB image of DECaLS, in the upper right panel, the RGB image of DECaLS without the detections and, in the lower right panel, we show the apertures image of RUN\,3 when executed on the DECaLS frames. Going back to Figure\,\ref{fig:RUN3-LEGACY}, the x-axis in the right panel shows the magnitudes measured by RUN\,3, and there the sources display the expected brightnesses. This is consistent with the goal of RUN\,3 of detecting and characterizing only bright and extended objects.

\subsection{Runs on more distant galaxy clusters: Antlia and Hydra}
\label{subsec:hydra_antlia}

As the over-deblending problem and the deficient detection of faint sources around bright galaxies was detected in S-PLUS images of the Fornax cluster \citep[D $\sim$ 20 Mpc,][]{Drinkwater2001}, it is interesting to check if those issues are also found in images of more distant clusters like Antlia \citep[D $\sim$ 35 Mpc,][]{Antlia2003A&A...408..929D} and Hydra (D $\sim$ 70 Mpc, \citealt{Hydra2012A&A...545A..37A}). In that context, we tested the performance of the iDR4, RUN\,1, RUN\,2 and RUN\,3 on S-PLUS images of those galaxy clusters. 

\begin{figure}
\centering
\includegraphics[width=1.0\columnwidth]{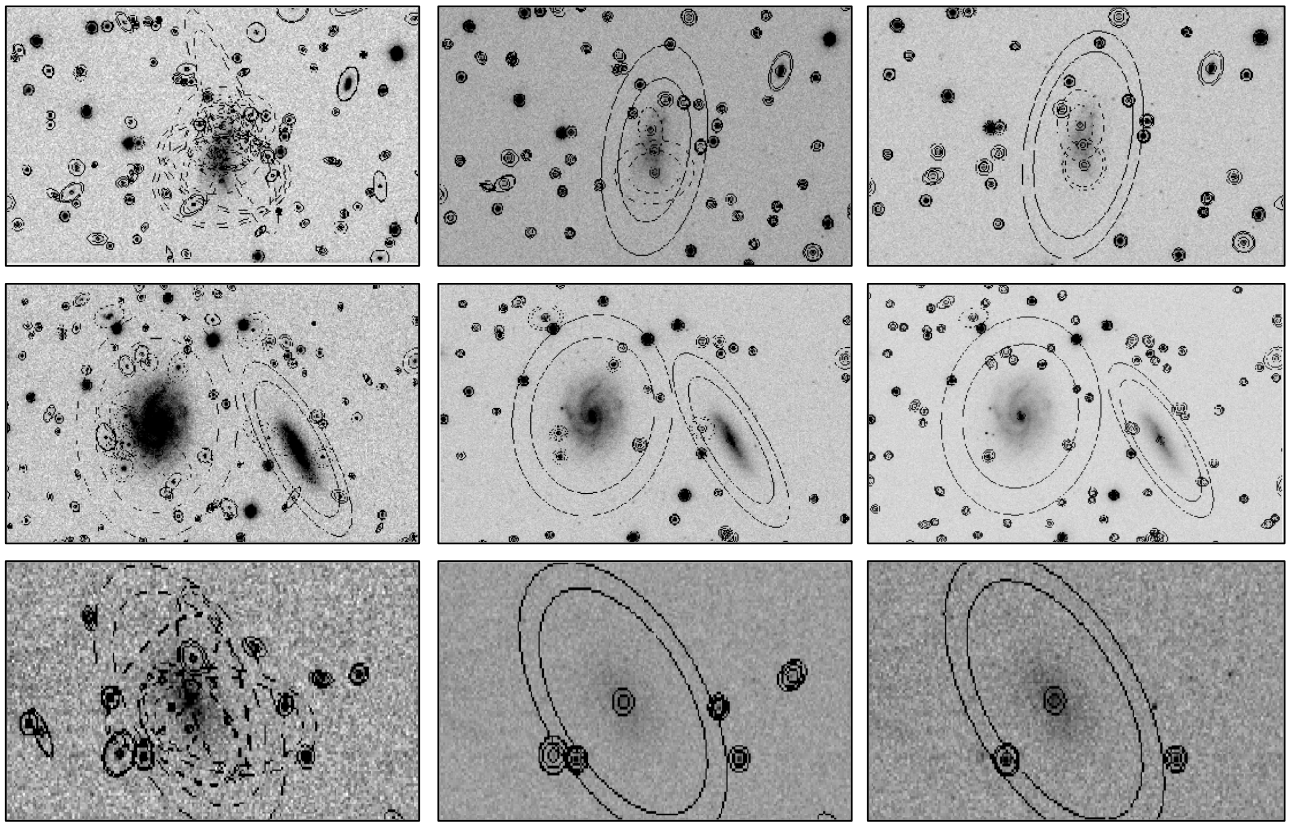}
 \caption{Hydra and Antlia examples. {\it Left:} Aperture images obtained using \texttt{SExtractor} input parameters of S-PLUS iDR4. {\it Center:} Aperture images obtained using RUN\,1. {\it Right:} Aperture images obtained from RUN\,2. All panels in a row show the same object.} 
 \label{fig:Hydra_Antlia}
\end{figure}

As it can be seen in Figure\,\ref{fig:Hydra_Antlia}, for Antlia and Hydra the iDR4 parameters still fail in the aperture setting, while RUN\,1, RUN\,2 and RUN\,3 reproduce the same good performance as on the Fornax cluster. Adding to the 106 S+FP fields the 333 Antlia and Hydra fields, our \texttt{SExtractor} parameters have been tested on $\sim$ 27\% of the total iDR4 S-PLUS pointings, significantly helping to solve the problem that motivated the present work. 

\subsection{Runs on images of other surveys}
\label{subsec:surveys}

\begin{figure*}
    \centering    
    \includegraphics[width=1.0\textwidth]{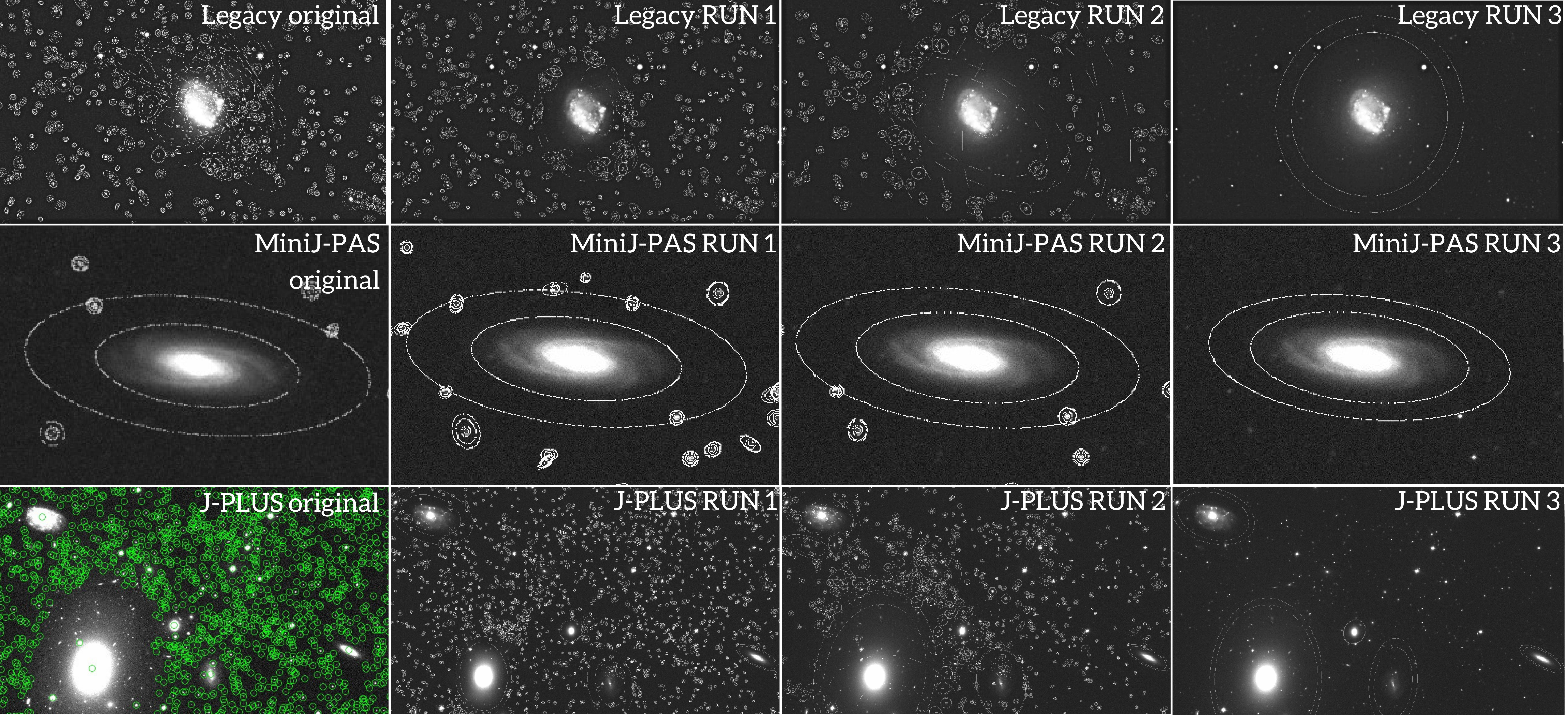}
 \caption{Twelve aperture images showing the performance of several sets of \texttt{SExtractor} parameters on images of different photometric surveys.  From top to bottom, each row corresponds to  DECaLS, MiniJ-PAS and J-PLUS, respectively. From left to right, the panels show the aperture images obtained by using the original \texttt{SExtractor} parameters of the survey, RUN\,1, RUN\,2 and RUN\,3, respectively. In the case of J-PLUS, for which we have not been able to find the set of parameters used to obtain the released catalogues, we depict in green the location of the objects in the catalogue downloaded from the J-PLUS database.}
\label{fig:compracion_surveys}
\end{figure*}

Given the results presented in the previous sections regarding DECaLS showing similar detection problems as S-PLUS, we wonder if over-deblending or missing objects also occur in the Javalambre Photometric Local Universe Survey \citep[J-PLUS,][]{J-PLUS} and in the publicly available data of the Javalambre Physics of the Accelerating Universe Astrophysical Survey (MiniJ-PAS, \citealp{Bonoli2021}). In addition, we would like to check the performance of RUN\,1, RUN\,2 and RUN\,3 on images of the aformentioned surveys. To that aim, in Figure\,\ref{fig:compracion_surveys} we show twelve aperture images where each row corresponds to a specific survey, and each column to the aperture images obtained using (from left to right) the original survey parameters, RUN\,1, RUN\,2 and RUN\,3, respectively. It is worth mentioning that in the case of J-PLUS we were not able to find the survey \texttt{SExtractor} parameters. In this case, the first panel shows the detections coming from the catalogue downloaded from the J-PLUS database. For DECaLS and MiniJ-PAS, the original parameters are given in Appendix\,\ref{sec:apen}.

As it can be seen in Figure\,\ref{fig:compracion_surveys}, the \texttt{SExtractor} parameters adopted by J-PLUS fail to properly detect very extended or small and faint extragalactic objects near bright ones, as S-PLUS and DECaLS. 
In the case of the parameters adopted by MiniJ-PAS, they seem to display a similar performance to RUN\,1 and RUN\,2. However, it will be interesting to analyze Javalambre Physics of the Accelerating Universe Astrophysical Survey \citep[J-PAS,][]{J-PAS} images of more extended objects than the ones analyzed in this work, once those images are made publicly available.

\section{Discussion}
\label{sec:discussion}

From our results it is clear that caution is required when using photometric catalogues coming from large photometric surveys that aim to address galactic and extragalactic topics. As a compromise is needed at choosing the most suitable \texttt{SExtractor} parameters to generate catalogues useful for both purposes, it may be possible that specific types of objects are not well characterized by a general configuration of those parameters.

As it was shown here, at distances lower than $D\sim50$ Mpc, commonly used \texttt{SExtractor} parameters may produce an over-deblending of star-forming galaxies and of large galaxies displaying complex morphologies. In addition, the lack of detection of faint and compact objects next to bright galaxies becomes a likely issue. From the analysis of four surveys like S-PLUS, J-PLUS, MiniJ-PAS and DECaLS, although in a reduced number of cases, three of them (75\%) display the above described situations. In this context, the implementation of alternative or parallel \texttt{SExtractor} runs that are focused on specific types of astrophysical objects turns out to be mandatory in order to be safe at the stage of making statistical studies. Such {\it Value Added Catalogues} (VACs) will represent additional and valuable outcomes on their own. 

In the specific case of S-PLUS, it is worth mentioning that the S+FP RUN\,1+2 catalogue is being used to select the spectroscopic targets in the Fornax cluster region for the CHANCES project within 4MOST \citep{CHANCES}. In addition, RUN\,1, RUN\,2 and RUN\,3 have been internally implemented on 27\% of the pointings of the S-PLUS iDR4, covering a total area of $\sim860$ deg$^2$ of the southern sky. Such an effort will allow the S-PLUS collaboration to start a comparison of the photometric properties of groups and clusters of galaxies at z<0.19, for which the H$\alpha$ line is within the S-PLUS H$\alpha$ filter. In the specific case of RUN\,1, it will allow the study of interesting compact objects like GCs and UCDs that were missed in the general catalogue of iDR4. 

Despite the specific issues found in the analysis of the galaxy content of nearby groups and clusters, we should emphasize that the catalogues of the now publicly available S-PLUS DR4\footnote{\url{https://splus.cloud/documentation/DR4}} are a valuable source for a number of studies. As it was also shown here, galaxies with smooth morphologies and not placed in the vicinity of bright sources are well characterized in those catalogues. Furthermore, none of the RUN\,1, RUN\,2 or RUN\,3 brings a particular improvement in the detection or measurement of that kind of objects. In that sense, it is worth mentioning that the catalogues of the S-PLUS DR4 contain confident and homogeneous 12-bands photometry for resolved and unresolved sources located in an area of $\sim$3,193 deg$^2$ of the southern sky, including the region known as Stripe-82 which is common to the Sloan Digital Sky Survey \citep[SDSS,][]{York2000}.

\section{Conclusions}
\label{sec:conclusions}

In this paper we present the identification of over-deblending issues in star-forming galaxies and an inefficient detection of faint objects near bright galaxies at the Fornax distance in the catalogues of S-PLUS iDR4. These issues arise as a result of the compromise reached in the definition of the \texttt{SExtractor} parameters chosen to obtain the photometric catalogues intended to be useful for both galactic and extragalactic studies. In addition, we found that those issues are also present in the catalogues recently released by other surveys.

In that context, we explore possible solutions testing several sets of \texttt{SExtractor} parameters on images from different multi-band surveys such as DECaLS, J-PLUS and MiniJPAS. It is important to mention that in this work we browse and analyze the differences in the \texttt{SExtractor} configuration files, mainly focusing on the parameters concerning the deblending, size and sensitivity of the source detection. 
As a future work, we plan to extend and deepen this analysis, trying to quantify other factors that contribute to the differences in the performance of the tested \texttt{SExtractor} runs. 

The results of our work can be summarized as follows:

\begin{itemize}
    \item S-PLUS iDR4 \texttt{SExtractor} parameters are useful to automatically detect and measure galaxies displaying smooth morphologies and that are not placed in the vicinity of bright objects.
    \item S-PLUS iDR4 \texttt{SExtractor} parameters miss faint galaxies and compact objects near bright galaxies at the Fornax cluster distance.
    \item S-PLUS iDR4 \texttt{SExtractor} parameters over-deblend star-forming galaxies at the Fornax, Antlia and Hydra clusters distances. 
    \item RUN\,1 \texttt{SExtractor} parameters recover faint objects near bright galaxies at the Fornax distance.
    \item RUN\,2 \texttt{SExtractor} parameters avoid the over-deblending of low and intermediate luminosity star-forming galaxies and make a good characterization of the sizes of large early-type galaxies.
    \item RUN\,2 \texttt{SExtractor} parameters do not avoid over-deblending in large star-forming galaxies. 
    \item RUN\,3 \texttt{SExtractor} parameters avoid the over-deblending of large star-forming galaxies and improve the characterization of the sizes of large early-type galaxies in comparison with RUN\,2.
    \item Strong over-deblending of star-forming galaxies is also present in DECaLS and J-PLUS catalogues. 
    \item RUN\,1, RUN\,2 and RUN\,3 display the same performance as on S-PLUS images when run on DECaLS, J-PLUS and MiniJPAS images. 
\end{itemize}

The main conclusion of this work is that catalogues coming from large photometric surveys that have both galactic and extragalactic goals should be used with caution, at least in nearby environments. 


\section*{Acknowledgements}
We are grateful to the referee, Dr. Florence Durret, for her detailed revision and useful analysis of our paper that greatly helped to improve its content. RFH, AVSC, FRF and ARL acknowledge financial support from CONICET, Agencia I+D+i (PICT 2019-03299) and Universidad Nacional de La Plata (Argentina). R.D. gratefully acknowledges support by the ANID BASAL project FB210003.

The S-PLUS project, including the T80-South robotic telescope and the S-PLUS scientific survey, was founded as a partnership between the Fundação de Amparo à Pesquisa do Estado de São Paulo (FAPESP), the Observatório Nacional (ON), the Federal University of Sergipe (UFS), and the Federal University of Santa Catarina (UFSC), with important financial and practical contributions from other collaborating institutes in Brazil, Chile (Universidad de La Serena), and Spain (Centro de Estudios de Física del Cosmos de Aragón, CEFCA). We further acknowledge financial support from the São Paulo Research Foundation (FAPESP), Fundação de Amparo à Pesquisa do Estado do RS (FAPERGS), the Brazilian National Research Council (CNPq), the Coordination for the Improvement of Higher Education Personnel (CAPES), the Carlos Chagas Filho Rio de Janeiro State Research Foundation (FAPERJ), and the Brazilian Innovation Agency (FINEP). The authors who are members of the S-PLUS collaboration are grateful for the contributions from CTIO staff in helping in the construction, commissioning and maintenance of the T80-South telescope and camera. We are also indebted to Rene Laporte and INPE, as well as Keith Taylor, for their important contributions to the project. From CEFCA, we particularly would like to thank Antonio Marín-Franch for his invaluable contributions in the early phases of the project, David Cristóbal-Hornillos and his team for their help with the installation of the data reduction package jype version 0.9.9, César Íñiguez for providing 2D measurements of the filter transmissions, and all other staff members for their support with various aspects of the project. R.D. gratefully acknowledges support by the ANID BASAL project FB210003.

\section*{Data Availability}

The data underlying this article will be shared on reasonable request to the corresponding author.




\bibliographystyle{mnras}
\bibliography{Haack} 

\begin{thebibliography}{}
\makeatletter
\relax
\def\mn@urlcharsother{\let\do\@makeother \do\$\do\&\do\#\do\^\do\_\do\%\do\~}
\def\mn@doi{\begingroup\mn@urlcharsother \@ifnextchar [ {\mn@doi@}
  {\mn@doi@[]}}
\def\mn@doi@[#1]#2{\def\@tempa{#1}\ifx\@tempa\@empty \href
  {http://dx.doi.org/#2} {doi:#2}\else \href {http://dx.doi.org/#2} {#1}\fi
  \endgroup}
\def\mn@eprint#1#2{\mn@eprint@#1:#2::\@nil}
\def\mn@eprint@arXiv#1{\href {http://arxiv.org/abs/#1} {{\tt arXiv:#1}}}
\def\mn@eprint@dblp#1{\href {http://dblp.uni-trier.de/rec/bibtex/#1.xml}
  {dblp:#1}}
\def\mn@eprint@#1:#2:#3:#4\@nil{\def\@tempa {#1}\def\@tempb {#2}\def\@tempc
  {#3}\ifx \@tempc \@empty \let \@tempc \@tempb \let \@tempb \@tempa \fi \ifx
  \@tempb \@empty \def\@tempb {arXiv}\fi \@ifundefined
  {mn@eprint@\@tempb}{\@tempb:\@tempc}{\expandafter \expandafter \csname
  mn@eprint@\@tempb\endcsname \expandafter{\@tempc}}}

\bibitem[\protect\citeauthoryear{{Almeida-Fernandes}
  et~al.,}{{Almeida-Fernandes} et~al.}{2022}]{Almeida-Fernandes2022}
{Almeida-Fernandes} F.,  et~al., 2022, \mn@doi [\mnras]
  {10.1093/mnras/stac284}, \href
  {https://ui.adsabs.harvard.edu/abs/2022MNRAS.511.4590A} {511, 4590}

\bibitem[\protect\citeauthoryear{{Ann}, {Seo}  \& {Ha}}{{Ann}
  et~al.}{2015}]{Ann2015}
{Ann} H.~B.,  {Seo} M.,   {Ha} D.~K.,  2015, \mn@doi [\apjs]
  {10.1088/0067-0049/217/2/27}, \href
  {https://ui.adsabs.harvard.edu/abs/2015ApJS..217...27A} {217, 27}

\bibitem[\protect\citeauthoryear{{Arnaboldi}, {Ventimiglia}, {Iodice},
  {Gerhard}  \& {Coccato}}{{Arnaboldi} et~al.}{2012}]{Hydra2012A&A...545A..37A}
{Arnaboldi} M.,  {Ventimiglia} G.,  {Iodice} E.,  {Gerhard} O.,   {Coccato} L.,
   2012, \mn@doi [\aap] {10.1051/0004-6361/201116752}, \href
  {https://ui.adsabs.harvard.edu/abs/2012A&A...545A..37A} {545, A37}

\bibitem[\protect\citeauthoryear{{Axelrod}}{{Axelrod}}{2006}]{Axelrod2006}
{Axelrod} T.~S.,  2006, in {Gabriel} C.,  {Arviset} C.,  {Ponz} D.,   {Enrique}
  S.,  eds,  Astronomical Society of the Pacific Conference Series Vol. 351,
  Astronomical Data Analysis Software and Systems XV. p.~103

\bibitem[\protect\citeauthoryear{{Barbosa} et~al.,}{{Barbosa}
  et~al.}{2020}]{Barbosa2020}
{Barbosa} C.~E.,  et~al., 2020, \mn@doi [\apjs] {10.3847/1538-4365/ab7660},
  \href {https://ui.adsabs.harvard.edu/abs/2020ApJS..247...46B} {247, 46}

\bibitem[\protect\citeauthoryear{{Benitez} et~al.,}{{Benitez}
  et~al.}{2014}]{J-PAS}
{Benitez} N.,  et~al., 2014, \mn@doi [arXiv e-prints]
  {10.48550/arXiv.1403.5237}, \href
  {https://ui.adsabs.harvard.edu/abs/2014arXiv1403.5237B} {p. arXiv:1403.5237}

\bibitem[\protect\citeauthoryear{{Bertin} \& {Arnouts}}{{Bertin} \&
  {Arnouts}}{1996}]{SExtractor}
{Bertin} E.,  {Arnouts} S.,  1996, \mn@doi [\aaps] {10.1051/aas:1996164}, \href
  {https://ui.adsabs.harvard.edu/abs/1996A&AS..117..393B} {117, 393}

\bibitem[\protect\citeauthoryear{{Blakeslee} et~al.,}{{Blakeslee}
  et~al.}{2009}]{Blakeslee2009}
{Blakeslee} J.~P.,  et~al., 2009, \mn@doi [\apj] {10.1088/0004-637X/694/1/556},
  \href {https://ui.adsabs.harvard.edu/abs/2009ApJ...694..556B} {694, 556}

\bibitem[\protect\citeauthoryear{{Bom} et~al.,}{{Bom} et~al.}{2021}]{Bom2021}
{Bom} C.~R.,  et~al., 2021, \mn@doi [\mnras] {10.1093/mnras/stab1981}, \href
  {https://ui.adsabs.harvard.edu/abs/2021MNRAS.507.1937B} {507, 1937}

\bibitem[\protect\citeauthoryear{{Bonoli} et~al.,}{{Bonoli}
  et~al.}{2021}]{Bonoli2021}
{Bonoli} S.,  et~al., 2021, \mn@doi [\aap] {10.1051/0004-6361/202038841}, \href
  {https://ui.adsabs.harvard.edu/abs/2021A&A...653A..31B} {653, A31}

\bibitem[\protect\citeauthoryear{{Buzzo} et~al.,}{{Buzzo}
  et~al.}{2022}]{Buzzo2022}
{Buzzo} M.~L.,  et~al., 2022, \mn@doi [\mnras] {10.1093/mnras/stab3489}, \href
  {https://ui.adsabs.harvard.edu/abs/2022MNRAS.510.1383B} {510, 1383}

\bibitem[\protect\citeauthoryear{{Cenarro} et~al.,}{{Cenarro}
  et~al.}{2019}]{J-PLUS}
{Cenarro} A.~J.,  et~al., 2019, \mn@doi [\aap] {10.1051/0004-6361/201833036},
  \href {https://ui.adsabs.harvard.edu/abs/2019A&A...622A.176C} {622, A176}

\bibitem[\protect\citeauthoryear{{Costa-Duarte} et~al.,}{{Costa-Duarte}
  et~al.}{2019}]{Costa-Duarte2019}
{Costa-Duarte} M.~V.,  et~al., 2019, \mn@doi [arXiv e-prints]
  {10.48550/arXiv.1909.08626}, \href
  {https://ui.adsabs.harvard.edu/abs/2019arXiv190908626C} {p. arXiv:1909.08626}

\bibitem[\protect\citeauthoryear{{Dey} et~al.,}{{Dey} et~al.}{2019a}]{Dey2019}
{Dey} A.,  et~al., 2019a, \mn@doi [\aj] {10.3847/1538-3881/ab089d}, \href
  {https://ui.adsabs.harvard.edu/abs/2019AJ....157..168D} {157, 168}

\bibitem[\protect\citeauthoryear{{Dey} et~al.,}{{Dey} et~al.}{2019b}]{Legacy}
{Dey} A.,  et~al., 2019b, \mn@doi [\aj] {10.3847/1538-3881/ab089d}, \href
  {https://ui.adsabs.harvard.edu/abs/2019AJ....157..168D} {157, 168}

\bibitem[\protect\citeauthoryear{{Dirsch}, {Richtler}  \& {Bassino}}{{Dirsch}
  et~al.}{2003}]{Antlia2003A&A...408..929D}
{Dirsch} B.,  {Richtler} T.,   {Bassino} L.~P.,  2003, \mn@doi [\aap]
  {10.1051/0004-6361:20031027}, \href
  {https://ui.adsabs.harvard.edu/abs/2003A&A...408..929D} {408, 929}

\bibitem[\protect\citeauthoryear{{Drinkwater}, {Gregg}  \&
  {Colless}}{{Drinkwater} et~al.}{2001}]{Drinkwater2001}
{Drinkwater} M.~J.,  {Gregg} M.~D.,   {Colless} M.,  2001, \mn@doi [\apjl]
  {10.1086/319113}, \href
  {https://ui.adsabs.harvard.edu/abs/2001ApJ...548L.139D} {548, L139}

\bibitem[\protect\citeauthoryear{{Guti{\'e}rrez-Soto}
  et~al.,}{{Guti{\'e}rrez-Soto} et~al.}{2020}]{Gutierrez-Soto2020}
{Guti{\'e}rrez-Soto} L.~A.,  et~al., 2020, \mn@doi [\aap]
  {10.1051/0004-6361/201935700}, \href
  {https://ui.adsabs.harvard.edu/abs/2020A&A...633A.123G} {633, A123}

\bibitem[\protect\citeauthoryear{{Haines} et~al.,}{{Haines}
  et~al.}{2023}]{CHANCES}
{Haines} C.,  et~al., 2023, \mn@doi [The Messenger] {10.18727/0722-6691/5308},
  \href {https://ui.adsabs.harvard.edu/abs/2023Msngr.190...31H} {190, 31}

\bibitem[\protect\citeauthoryear{{Hartmann} et~al.,}{{Hartmann}
  et~al.}{2022}]{Hartmann2022}
{Hartmann} E.~A.,  et~al., 2022, \mn@doi [\mnras] {10.1093/mnras/stac1411},
  \href {https://ui.adsabs.harvard.edu/abs/2022MNRAS.515.4191H} {515, 4191}

\bibitem[\protect\citeauthoryear{{Lima-Dias} et~al.,}{{Lima-Dias}
  et~al.}{2021}]{Lima-Dias2021}
{Lima-Dias} C.,  et~al., 2021, \mn@doi [\mnras] {10.1093/mnras/staa3326}, \href
  {https://ui.adsabs.harvard.edu/abs/2021MNRAS.500.1323L} {500, 1323}

\bibitem[\protect\citeauthoryear{{Lima-Dias} et~al.,}{{Lima-Dias}
  et~al.}{2024}]{Lima-Dias2024}
{Lima-Dias} C.,  et~al., 2024, \mn@doi [\mnras] {10.1093/mnras/stad3571}, \href
  {https://ui.adsabs.harvard.edu/abs/2024MNRAS.527.5792L} {527, 5792}

\bibitem[\protect\citeauthoryear{{Lima} et~al.,}{{Lima}
  et~al.}{2022}]{Lima2022}
{Lima} E.~V.~R.,  et~al., 2022, \mn@doi [Astronomy and Computing]
  {10.1016/j.ascom.2021.100510}, \href
  {https://ui.adsabs.harvard.edu/abs/2022A&C....3800510L} {38, 100510}

\bibitem[\protect\citeauthoryear{{Lintott} et~al.,}{{Lintott}
  et~al.}{2011}]{GZ1}
{Lintott} C.,  et~al., 2011, \mn@doi [\mnras]
  {10.1111/j.1365-2966.2010.17432.x}, \href
  {https://ui.adsabs.harvard.edu/abs/2011MNRAS.410..166L} {410, 166}

\bibitem[\protect\citeauthoryear{{Mendes de Oliveira} et~al.,}{{Mendes de
  Oliveira} et~al.}{2019}]{Mendes2019}
{Mendes de Oliveira} C.,  et~al., 2019, \mn@doi [\mnras]
  {10.1093/mnras/stz1985}, \href
  {https://ui.adsabs.harvard.edu/abs/2019MNRAS.489..241M} {489, 241}

\bibitem[\protect\citeauthoryear{{Montaguth} et~al.,}{{Montaguth}
  et~al.}{2023}]{Montaguth2023}
{Montaguth} G.~P.,  et~al., 2023, \mn@doi [\mnras] {10.1093/mnras/stad2235},
  \href {https://ui.adsabs.harvard.edu/abs/2023MNRAS.524.5340M} {524, 5340}

\bibitem[\protect\citeauthoryear{{Nair} \& {Abraham}}{{Nair} \&
  {Abraham}}{2010}]{Nair2010}
{Nair} P.~B.,  {Abraham} R.~G.,  2010, \mn@doi [\apjs]
  {10.1088/0067-0049/186/2/427}, \href
  {https://ui.adsabs.harvard.edu/abs/2010ApJS..186..427N} {186, 427}

\bibitem[\protect\citeauthoryear{{Nakazono} et~al.,}{{Nakazono}
  et~al.}{2021}]{Nakazono2021}
{Nakazono} L.,  et~al., 2021, \mn@doi [\mnras] {10.1093/mnras/stab1835}, \href
  {https://ui.adsabs.harvard.edu/abs/2021MNRAS.507.5847N} {507, 5847}

\bibitem[\protect\citeauthoryear{{Oke}}{{Oke}}{1974}]{Oke1974ApJS...27...21O}
{Oke} J.~B.,  1974, \mn@doi [\apjs] {10.1086/190287}, \href
  {https://ui.adsabs.harvard.edu/abs/1974ApJS...27...21O} {27, 21}

\bibitem[\protect\citeauthoryear{{Simmons} et~al.,}{{Simmons}
  et~al.}{2017}]{Simmons2017}
{Simmons} B.~D.,  et~al., 2017, \mn@doi [\mnras] {10.1093/mnras/stw2587}, \href
  {https://ui.adsabs.harvard.edu/abs/2017MNRAS.464.4420S} {464, 4420}

\bibitem[\protect\citeauthoryear{{Smith Castelli} et~al.,}{{Smith Castelli}
  et~al.}{2024}]{PaperI}
{Smith Castelli} A.~V.,  et~al., 2024, \mn@doi [\mnras]
  {10.1093/mnras/stae840}, \href
  {https://ui.adsabs.harvard.edu/abs/2024MNRAS.tmp..868C} {}

\bibitem[\protect\citeauthoryear{{Spiekermann}}{{Spiekermann}}{1992}]{Spiekermann1992}
{Spiekermann} G.,  1992, \mn@doi [\aj] {10.1086/116215}, \href
  {https://ui.adsabs.harvard.edu/abs/1992AJ....103.2102S} {103, 2102}

\bibitem[\protect\citeauthoryear{{Storrie-Lombardi}, {Lahav}, {Sodre}  \&
  {Storrie-Lombardi}}{{Storrie-Lombardi} et~al.}{1992}]{Storrie-Lombardi1992}
{Storrie-Lombardi} M.~C.,  {Lahav} O.,  {Sodre} L. J.,   {Storrie-Lombardi}
  L.~J.,  1992, \mn@doi [\mnras] {10.1093/mnras/259.1.8P}, \href
  {https://ui.adsabs.harvard.edu/abs/1992MNRAS.259P...8S} {259, 8P}

\bibitem[\protect\citeauthoryear{{Thain{\'a}-Batista}
  et~al.,}{{Thain{\'a}-Batista} et~al.}{2023}]{Thaina-Batista2023}
{Thain{\'a}-Batista} J.,  et~al., 2023, \mn@doi [\mnras]
  {10.1093/mnras/stad2698}, \href
  {https://ui.adsabs.harvard.edu/abs/2023MNRAS.tmp.2606T} {}

\bibitem[\protect\citeauthoryear{{Tyson}, {Wittman}, {Hennawi}  \&
  {Spergel}}{{Tyson} et~al.}{2002}]{Tyson2002}
{Tyson} T.,  {Wittman} D.,  {Hennawi} J.,   {Spergel} D.,  2002, in APS April
  Meeting Abstracts. APS Meeting Abstracts.
p. Y6.004

\bibitem[\protect\citeauthoryear{{Walmsley} et~al.,}{{Walmsley}
  et~al.}{2020}]{Walmsley2020}
{Walmsley} M.,  et~al., 2020, \mn@doi [\mnras] {10.1093/mnras/stz2816}, \href
  {https://ui.adsabs.harvard.edu/abs/2020MNRAS.491.1554W} {491, 1554}

\bibitem[\protect\citeauthoryear{{Werner} et~al.,}{{Werner}
  et~al.}{2023}]{Werner2023}
{Werner} S.~V.,  et~al., 2023, \mn@doi [\mnras] {10.1093/mnras/stac3273}, \href
  {https://ui.adsabs.harvard.edu/abs/2023MNRAS.519.2630W} {519, 2630}

\bibitem[\protect\citeauthoryear{{Willett} et~al.,}{{Willett}
  et~al.}{2013}]{GZ2}
{Willett} K.~W.,  et~al., 2013, \mn@doi [\mnras] {10.1093/mnras/stt1458}, \href
  {https://ui.adsabs.harvard.edu/abs/2013MNRAS.435.2835W} {435, 2835}

\bibitem[\protect\citeauthoryear{{York} et~al.,}{{York}
  et~al.}{2000}]{York2000}
{York} D.~G.,  et~al., 2000, \mn@doi [\aj] {10.1086/301513}, \href
  {https://ui.adsabs.harvard.edu/abs/2000AJ....120.1579Y} {120, 1579}

\makeatother
\end{thebibliography}



\newpage
\appendix

\section{Photometric Depth}
\label{sec:apen2}

\begin{figure}
        \centering        
        \includegraphics[width=1.0\columnwidth]{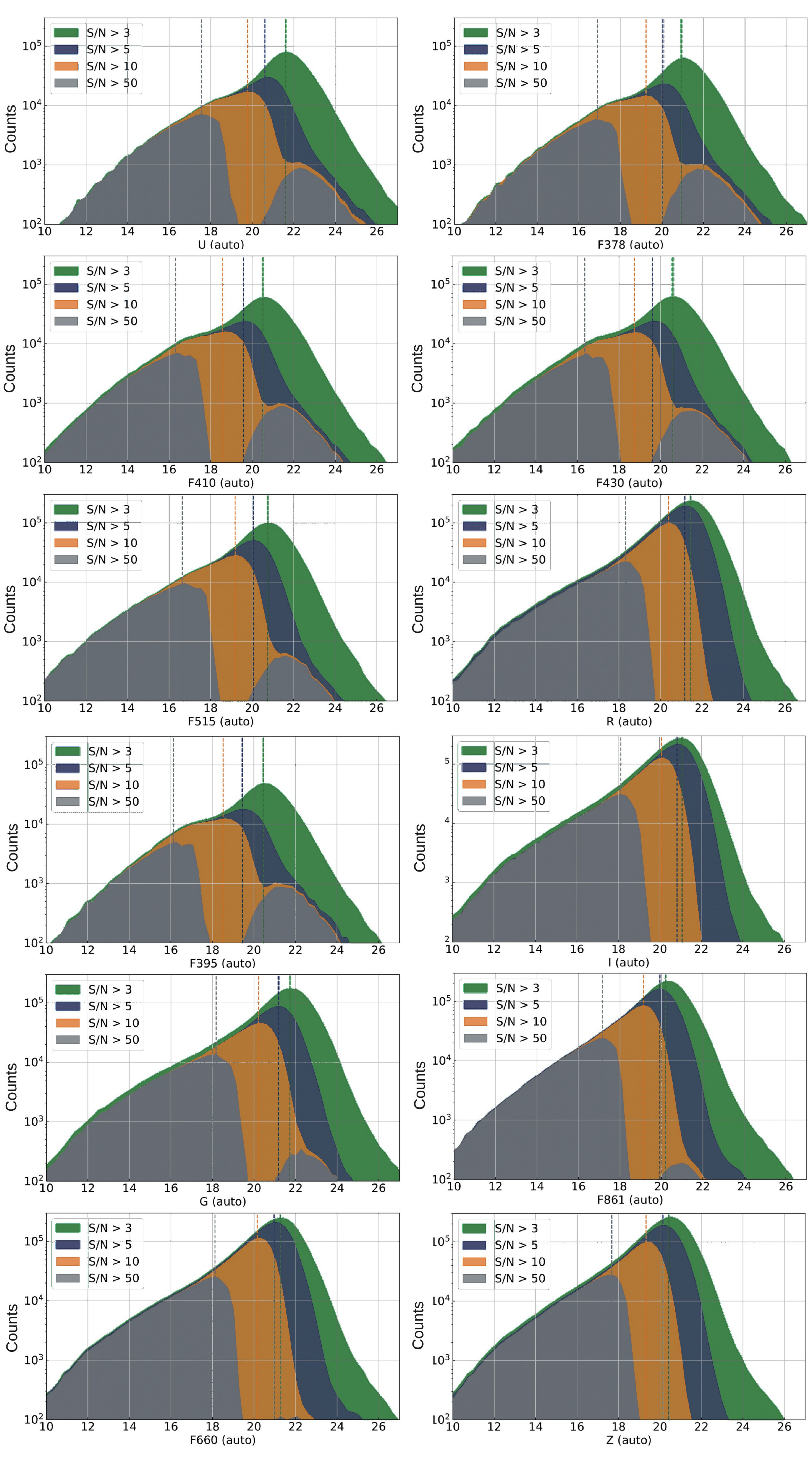}
\caption{Photometric depth of RUN\,1 in the 12 bands of S-PLUS considering four signal-to-noise (S/N) thresholds (S/N > 3, green; S/N > 5, blue; S/N > 10, orange; S/N > 50, grey) for each filter \citep{Almeida-Fernandes2022}. The characteristic photometric depth at each S/N threshold and for each filter is given by the location of the peak of the magnitude distribution (dashed coloured lines).} 
\label{fig:completitud}
\end{figure}

An interesting analysis to perform is to estimate the depth of the S+FP photometry in each of the twelve filters of S-PLUS. It makes sense to do this using the catalogue obtained with RUN\,1, since it is the one that recovers the faintest objects. In Figure\,\ref{fig:completitud}, we show depth plots (counts vs. mag AUTO) for each of the filters considering the same S/N ranges as \citet{Almeida-Fernandes2022} (S/N > 3, S/N > 5, S/N > 10 and S/N > 50). The depth value is the magnitude at the peak (marked by the vertical lines) of each of the distributions. It can be seen that in some filters, and for S/N > 10, a second fainter peak appears. These secondary peaks correspond to spurious detections arising from bad pixels, very noisy sky regions and objects near the edges of the frames, among other artifacts.

\begin{figure*}
    \centering
    \includegraphics[width=1.0\textwidth]{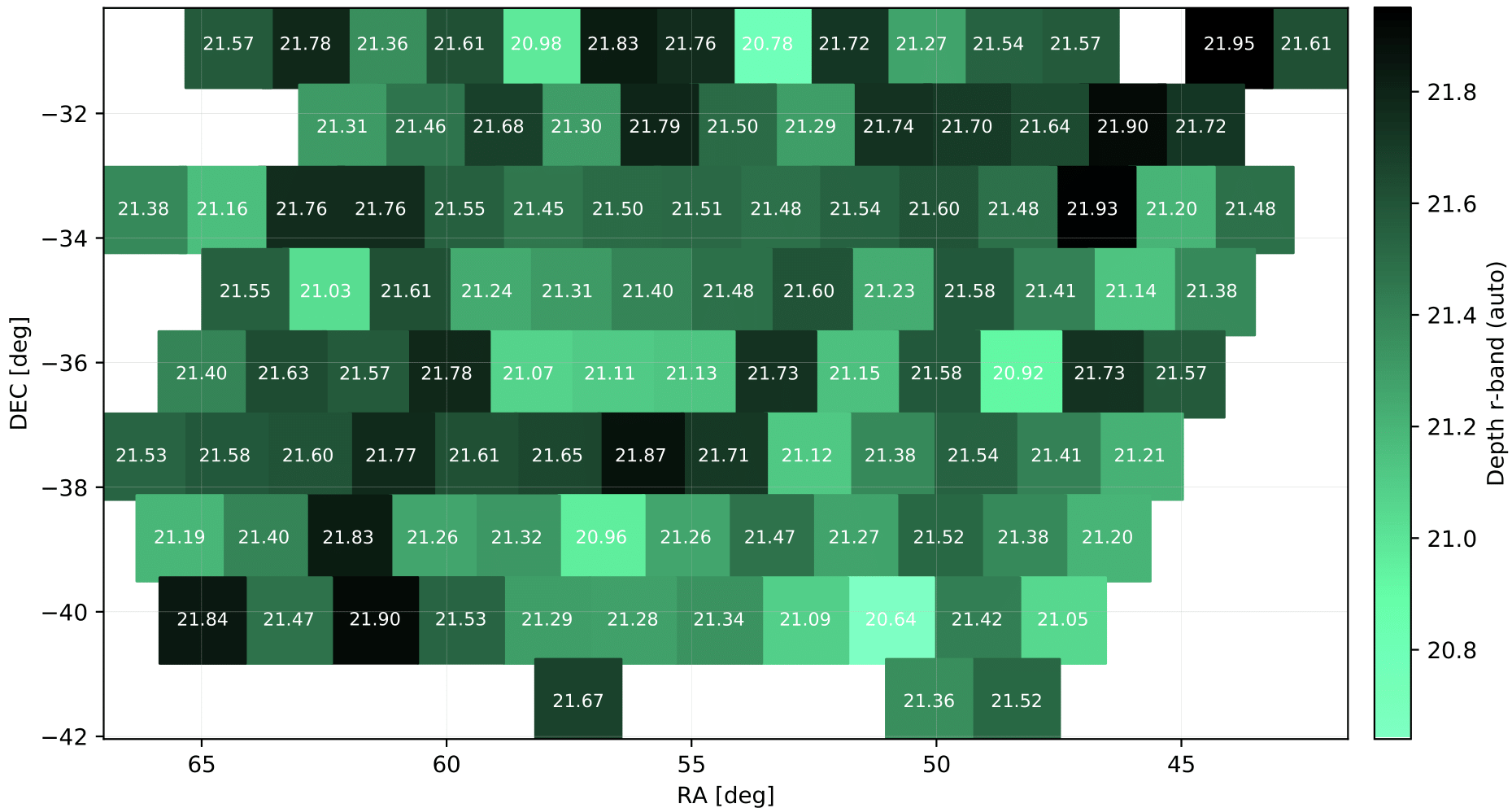}
 \caption{Photometric depth of the 106 S+FP fields in the r-band. The values shown in this plot correspond to the AUTO r-band magnitude of the faintest object detected in each field by RUN\,1 (considering sources with S/N > 3), which allows us to detect the faintest and most compact objects of the whole S+FP photometric sample.}
\label{Depth}
\end{figure*}

It is also possible to map the depth corresponding to each of the 106 fields of the S+FP. Considering only objects with S/N > 3, Figure\,\ref{Depth} shows the depth values in the r-band. This depth map allow us to see, in a simple way, not only how the depth changes from field to field but also that, within the considered S-PLUS pointings, there are different subregions with a diversity in photometric depth. 

The estimates from Figure\,\ref{fig:completitud} and Figure\,\ref{Depth} allow us to compare the depth values obtained from RUN\,1 with those arising from S-PLUS iDR4 \citep{Almeida-Fernandes2022}. In Table\,\ref{Tabla_depth}, we present the limiting values corresponding to the peaks in Figure\,\ref{fig:completitud} for each filter and each S/N value. It can be seen that the RUN\,1 photometry is slightly deeper than that of S-PLUS iDR4. In addition, to complete the description of Figure\,\ref{Depth}, the statistical parameters obtained for the 106 S+FP fields are shown in the Table \ref{tab:Statistics}.

\begin{table}
\centering
\caption{Depth limit for the 12 filters at different S/N values, considering the RUN\,1 catalogue that includes the faintest objects of both S+FP \texttt{SExtractor} runs.}
\begin{tabular}{c|c|c|c|c}
\textbf{Filter} & \textbf{S/N > 3} & \textbf{S/N > 5} & \textbf{S/N > 10} & \textbf{S/N > 50} \\
\hline
u      & 21.36              & 20.43              & 19.50              & 17.17              \\
g      & 21.61              & 21.15              & 20.21              & 17.88              \\
r      & 21.37              & 20.91              & 19.98              & 18.12              \\
i      & 20.96              & 20.50              & 20.03              & 17.70              \\
z      & 20.48              & 20.01              & 19.07              & 17.19              \\
F0378   & 20.59              & 19.66              & 19.19              & 16.86              \\
F0395   & 20.47              & 19.06              & 18.60              & 15.79              \\
F0410   & 20.44              & 19.50              & 18.57              & 16.23              \\
F0430   & 20.47              & 19.54              & 18.60              & 16.26              \\
F0515   & 20.63              & 19.70              & 18.77              & 16.43              \\
F0660   & 21.25              & 20.79              & 19.86              & 17.99              \\
F0861   & 20.01              & 19.54              & 19.07              & 16.72              \\
\bottomrule
\label{Tabla_depth}
\end{tabular}
\end{table}

\begin{table*}
\centering
\caption{Statistical parameters obtained for the 106 S+FP fields in the $r$-band, considering sources with S/N > 3 in RUN\,1.}
\label{tab:Statistics}
\begin{tabular}{|c|c|c|c|c|c|c|c|}
\hline
Mean & Median & SD & Q$_{10}$ & Quartile$_1$ & Quartile$_3$ & Q$_{90}$ \\
\hline
21.37 mag & 21.4 mag & 0.2578 & 21.02 mag & 21.19 mag & 21.53 mag & 21.68 mag \\
\hline
\end{tabular}
\end{table*}

\section{SExtractor configuration files}
\label{sec:apen}

In this section we present the set of \texttt{SExtractor} parameters used by other photometric surveys.

\begin{table}[h]
\centering
\caption{List of SExtractor’s parameters used to build the miniJPAS source catalogues. Table C.1 from \citep{Bonoli2021}}
\label{tab:Parameters MiniJPAS}
\begin{tabular}{|l|r|}
\textbf{SExtractor Parameter} & \textbf{Value} \\
\hline
ANALYSIS\_THRESH & 2.0 \\
BACKPHOTO\_THICK & 24 \\
BACKPHOTO\_TYPE & LOCAL \\
BACK\_FILTERSIZE & 3 \\
BACK\_FILTTHRESH & 0.0 \\
BACK\_SIZE & 512 \\
BACK\_TYPE & AUTO \\
BACK\_VALUE & 0.0, 0.0 \\
CLEAN & Y \\
CLEAN\_PARAM & 1.0 \\
DEBLEND\_MINCONT & 0.005 \\
DEBLEND\_NTHRESH & 32 \\
DETECT\_MINAREA & 5 \\
DETECT\_THRESH & 0.9 (dual-mode) / 2 (single-mode) \\
DETECT\_TYPE & CCD \\
FILTER & Y \\
FILTER\_NAME & gauss\_3.0\_5x5.conv \\
GAIN\_KEY & GAIN \\
MASK\_TYPE & CORRECT \\
PHOT\_AUTOAPERS & 0.0, 0.0 \\
PHOT\_AUTOPARAMS & 2.5, 3.5 \\
PHOT\_FLUXFRAC & 0.5 \\
PHOT\_PETROPARAMS & 2.0, 3.5 \\
PIXEL\_SCALE & 0.2267 \\
STARNNW\_NAME & default.nnw \\
THRESH\_TYPE & RELATIVE \\
VERBOSE\_TYPE & NORMAL \\
WEIGHT\_GAIN & Y \\
WEIGHT\_THRESH & 0, 0 \\
WEIGHT\_TYPE & MAP\_WEIGHT, MAP\_WEIGHT \\
\end{tabular}
\end{table}

\begin{table}[h]
\centering
\caption{List of SExtractor’s parameters used by DECaLS \citep{Legacy}.} 
\label{tab:SExtractorDECaLS}
\begin{tabular}{|l|r|}
\textbf{SExtractor Parameter} & \textbf{Value} \\
\hline
ANALYSIS\_THRESH & 1.0 \\
BACK\_SIZE & 64 \\
BACK\_FILTERSIZE & 3 \\
BACKPHOTO\_TYPE & LOCAL \\
BACKPHOTO\_THICK & 24 \\
CATALOG\_TYPE & FITS\_LDAC \\
CLEAN & Y \\
CLEAN\_PARAM & 1.0 \\
CHECKIMAGE\_TYPE & NONE \\
DEBLEND\_NTHRESH & 64 \\
DEBLEND\_MINCONT & 0.000015 \\
DETECT\_THRESH & 1.0 \\
DETECT\_TYPE & CCD \\
DETECT\_MINAREA & 5 \\
DETECT\_MAXAREA & 0 \\
FILTER & Y \\
FILTER\_NAME & gauss\_5.0\_9x9.conv \\
GAIN & 4.0 \\
MAG\_ZEROPOINT & 30.0 \\
MASK\_TYPE & CORRECT \\
MEMORY\_OBJSTACK & 30000 \\
MEMORY\_PIXSTACK & 10000000 \\
MEMORY\_BUFSIZE & 1024 \\
PHOT\_APERTURES & 5.7251911, 11.450382, \\
PHOT\_AUTOPARAMS & 2.5, 3.5 \\
PHOT\_PETROPARAMS & 2.0, 3.5 \\
PIXEL\_SCALE & 0 \\
SATUR\_LEVEL & 40000.0 \\
SEEING\_FWHM & 1.2 \\
THRESH\_TYPE & RELATIVE \\
FITS\_UNSIGNED & Y \\
&  19.083969, 26.717558, \\
&  34.351147, 41.984734, \\
&  49.618320, 57.251911 \\
VERBOSE\_TYPE & QUIET \\
WRITE\_XML & N \\
\end{tabular}
\end{table}



\bsp	
\label{lastpage}
\end{document}